\documentclass[usenatbib]{mn2e}

\pdfoutput=1

\usepackage{fixltx2e}
\usepackage{mathptmx}
\usepackage{color}
\usepackage[pdftex]{graphicx}

\newcommand{\Mpc}{\rm\thinspace Mpc}
\newcommand{\kpc}{\rm\thinspace kpc}

\newcommand{\km}{\rm\thinspace km}

\newcommand{\cm}{\rm\thinspace cm}

\newcommand{\cmpssq}{\hbox{$\cm\s^{-2}\,$}}

\newcommand{\pcmcu}{\hbox{$\cm^{-3}\,$}}

\newcommand{\yr}{\rm\thinspace yr}

\newcommand{\s}{\rm\thinspace s}

\newcommand{\Hz}{\rm\thinspace Hz}

\newcommand{\keVpcmcu}{\hbox{$\keV\cm^{-3}\,$}}

\newcommand{\Msun}{\hbox{$\rm\thinspace M_{\odot}$}}

\newcommand{\Msunpyr}{\hbox{$\Msun\yr^{-1}\,$}}

\newcommand{\keV}{\rm\thinspace keV}

\newcommand{\erg}{\rm\thinspace erg}

\newcommand{\ergps}{\hbox{$\erg\s^{-1}\,$}}

\newcommand{\kmps}{\hbox{$\km\s^{-1}\,$}}

\newcommand{\kmpspMpc}{\hbox{$\kmps\Mpc^{-1}$}}

\newcommand{\Zsun}{\hbox{$\thinspace \mathrm{Z}_{\odot}$}}

\newcommand{\psqcm}{\hbox{$\cm^{-2}\,$}}

\begin{document}

\title{Giant cavities, cooling and metallicity substructure in
  Abell~2204} \author[J.~S. Sanders, A.~C. Fabian \& G.~B. Taylor]
{J.~S. Sanders$^1$\thanks{E-mail: jss@ast.cam.ac.uk},
  A.~C. Fabian$^1$ and G.~B. Taylor$^2$\\
  $^1$ Institute of Astronomy, Madingley Road, Cambridge. CB3 0HA\\
  $^2$ Department of Physics and Astronomy, University of New Mexico,
  Albuquerque, NM 87131. USA\\
}
\maketitle
  
\begin{abstract}
  We present results from deep \emph{Chandra} and \emph{XMM-Newton}
  observations of the relaxed X-ray luminous galaxy cluster
  Abell~2204. We detect metallicity inhomogeneities in the
  intracluster medium on a variety of distance scales, from a $\sim
  12$~kpc enhancement containing a few times $10^7 \Msun$ of iron in
  the centre, to a region at 400~kpc radius with an excess of a few
  times $10^9 \Msun$.  Subtracting an average surface brightness
  profile from the X-ray image yields two surface brightness
  depressions to the north and south of the cluster. Their morphology
  is similar to the cavities observed in cluster cores, but they have
  radii of 240~kpc and 160~kpc and have a total enthalpy of $2 \times
  10^{62}$~erg. If they are fossil radio bubbles, their buoyancy
  timescales imply a total mechanical heating power of $5 \times
  10^{46} \ergps$, the largest such bubble heating power known. More
  likely, they result from the accumulation of many past
  bubbles. Energetically this is more feasible, as the enthalpy of
  these regions could combat X-ray cooling in this cluster to 500~kpc
  radius for around 2~Gyr.  The core of the cluster also contains five
  to seven $\sim 4$~kpc radius surface brightness depressions that are
  not associated with the observed radio emission. If they are bubbles
  generated by the nucleus, they are too small to balance cooling in
  the core by an order of magnitude. However if the radio axis is
  close to the line of sight, projection effects may mask more normal
  bubbles. Using RGS spectra we detect a Fe~\textsc{xvii}
  line. Spectral fitting reveals temperatures down to $\sim 0.7$~keV;
  the cluster therefore shows a range in X-ray temperature of at least
  a factor of 15. The quantity of low temperature gas is consistent
  with a mass deposition rate of $65 \Msunpyr$.
\end{abstract}


\section{Introduction}
Abell~2204 is a luminous galaxy cluster at a redshift of 0.1523 ($L_X
= 2 \times 10^{45} h_{50}^{-2} \ergps$ in the 2-10 keV band;
\citealt{Edge90}).  On large scales the cluster has a regular
morphology \citep{BuoteTsai96,Schuecker01,Hashimoto07}, appearing
relaxed. \cite{Reiprich08} recently used a \emph{Suzaku} observation
of this cluster to measure its temperature near to the virial radius,
finding it close to predictions from hydrodynamic simulations.

We previously analysed a short snapshot observation of the cluster by
\emph{Chandra} \citep{SandersA220405}. The core of the cluster has a
fairly complex structure, containing a flat plateau and plume-like
feature. One of the most peculiar features was a high-metallicity ring
found around the core of the cluster, which we hypothesised may have
been due to a merger in the past.

Here we present results from a deep 77~ks \emph{Chandra} observations
of this cluster, in addition to new VLA radio observations. We assume
$H_0 = 70 \kmpspMpc$ and $\Omega_\Lambda = 0.7$, translating to a
scale of 2.6~kpc per arcsec. Metallicity measurements assume the
relative Solar abundances of \cite{AndersGrevesse89}. Abell~2204 has a
weighted Galactic hydrogen column density of $5.7\times 10^{20}
\psqcm$ determined by HI surveys \citep{Kalberla05}.

\section{Data processing}
The data analysed in this paper come from three different observations
of the cluster by \emph{Chandra} (Table~\ref{tab:observations}). Two
of the observations were taken in the ACIS-I detector mode, which uses
lower detector background front-illuminated CCDs. The other
observation was made which with the ACIS-S detector mode, which has a
higher background, but larger effective area, smaller field of view,
and better energy resolution.

\begin{table}
  \caption{\emph{Chandra} observations analysed in this paper.}
  \begin{tabular}{llll}
    Observation ID & Detector & Observation date & Exposure (ks) \\ \hline
    499 & ACIS-S & 2000-07-29 & 10.1 \\
    6104 & ACIS-I & 2004-09-20 & 9.6 \\
    7940 & ACIS-I & 2007-06-06 & 77.1 \\
  \end{tabular}
  \label{tab:observations}
\end{table}

For the part of the analysis concentrated in the bright central
region, we use all three datasets in combination. In the outer region
we concentrate on the 7940 and 6104 datasets, or just the 7940
dataset, as they have a lower non-X-ray background.

Each of the observations was made with the ACIS VFAINT mode. We
applied this VFAINT filtering to reduce the detector background, after
ensuring the observations used the latest gain files.

\begin{figure*}
  \includegraphics[width=\textwidth]{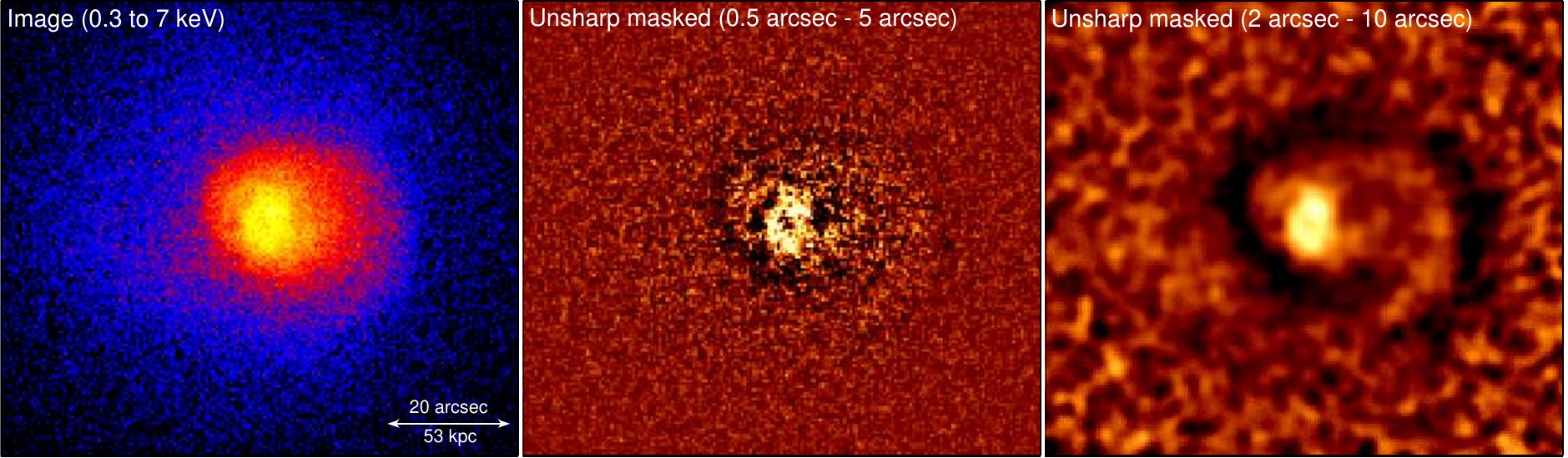}
  \caption{(Left panel) 0.3 to 7~keV image of the core the cluster
    with 0.492~arcsec pixels. (Centre panel) Unsharp-masked image,
    created by subtracting images smoothed by Gaussians of
    $\sigma=0.5$ and 5 pixels (regions with more emission on small
    scales are shown as lighter here). (Right panel) Unsharp-masked
    image, created by subtracting images smoothed with $\sigma=2$ and
    10~arcsec.}
  \label{fig:unsharp}
\end{figure*}

\subsection{Background modelling}
\label{sect:bg}
Comparison of spectra extracted from the edge of the ACIS-I
observations compared to standard blank-sky observations showed excess
soft emission. This soft emission appears to be spatially flat, not
declining with radius from the cluster centre, indicating it is not
cluster emission and likely Galactic in origin. The emission is very
similar between the 6104 and 7940 observations, showing it is not a
time-variable background.

Rather than use standard blank-sky observations, we decided to
construct our own backgrounds to account for this soft
emission. Firstly we took stowed background observations (where the
detector is not observing the sky). These observations, after
normalising to the observations in the 9-12 keV band and removing
VFAINT-filtered events, closely match the particle and detector
backgrounds.

To account for the Galactic and extragalactic X-ray emission, we
modelled a far off axis region (after accounting for the particle
background) with a thermal model (fixed at a temperature of 0.25~keV
and Solar metallicity) plus powerlaw ($\Gamma = 1.5$) model. This
model was a good fit to the data ($\chi^2_\nu = 211/225 =
0.94$). Taking this model, we simulated spectra in a grid over the
detector (iterating in $8\times8$ detector pixel cells) for each of
the observations, taking into account the change in effective area by
using ancillary response files generated at each grid point. By
deconstructing the spectra into individual events (randomising
spatially within each grid point), we generated simulated events files
for each observation. This X-ray background event file was then merged
with the normalised stowed background event file to make a total
background event file (we simulated the X-ray background file to have
the same exposure time as the stowed background). This part-synthetic
background provides a very good match to the spectra extracted from
the edge of detector and accounts for the particle background in the
centre of the observation properly.

The advantage of this procedure over modelling the soft component in
each spectral fit, is that we can easily account for detector
variation (bad pixels, vignetting, etc) and that it simplifies the
spectral fitting by only having one background dataset per
observation.

\begin{table*}
  \caption{Radio observational parameters.}
  \begin{center}
    \begin{tabular}{l l r r r r r r r r}
      Source & Date & Frequency & Bandwidth & Config. & Duration \\
      &  & (MHz) & (MHz) &  & (min) \\
      \hline
      \noalign{\vskip2pt}
      J1632+0534 & Jul 2007 & 322/329  & 6.25 & A & 171 \\
      & Nov 2007 & 1365/1435   & 50 & B & 83 \\
      & Apr 1998 & 1365/1435  & 25 & A & 240 \\
      & Apr 1998 & 4635/4885  & 50 & A & 61 \\
      & Aug 1998 & 8115/8485  & 50 & B & 134 \\
      \hline
    \end{tabular}
  \end{center}
  \label{table:radioparams}
\end{table*}

\subsection{Radio observations}
VLA observations of the radio source, J1632+0534 at the centre of
A2204, were performed on 2007 July 3 at 0.329 GHz in the ``A''
configuration, and on 2007 November 25 at 1.4 GHz in the ``B''
configuration.  We also make use of data from the VLA archive at 1.4,
5 and 8.4 GHz.  Details regarding the radio observations are
summarised in Table~\ref{table:radioparams}.  All data were reduced in
\textsc{aips} (Astronomical Image Processing System) following the
standard procedures.  Absolute flux density calibration was tied to
observations of 3C286.

\section{Imaging}
\subsection{Central structure}
In Fig.~\ref{fig:unsharp} (left panel) we show a merged
exposure-corrected image of the core of the cluster. As previously
described in \cite{SandersA220405}, there is a core with flat surface
brightness of dimensions of $7 \times 9$ arcsec. This core is embedded
within another flat central ``plateau'' of radius $\sim 10$~arcsec.  A
plume-like feature extends from the west of the plateau, wrapping
around from south to east.

The unsharp masked images in Fig.~\ref{fig:unsharp} (centre and right
panels) show there is considerable structure visible inside the inner
6~arcsec radius, and the surrounding plume.

\begin{figure*}
  \includegraphics[width=0.95\textwidth]{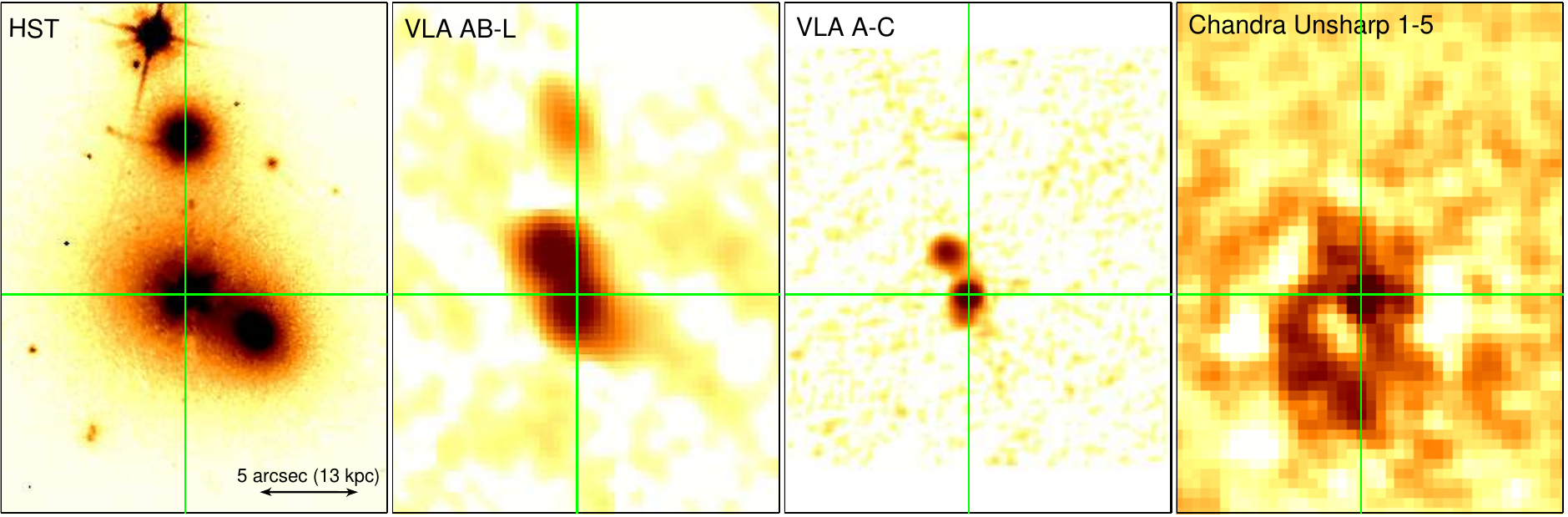}
  \caption{Comparison of \emph{HST} image of the central galaxies, VLA
    L band (20~cm) image, VLA C band (6~cm) image and \emph{Chandra}
    unsharp-masked image. The \emph{HST} image was created by
    combining datasets U5A44101R and U5A44102R (F606W filter).}
  \label{fig:comparebands}
\end{figure*}

The central radio nucleus (J1632+0534) corresponds with a peak in the
X-ray emission offset to the north of the centroid of the cluster
(Fig.~\ref{fig:comparebands}). The X-ray spectrum is compatible with a
$\Gamma=2.0 \pm 0.2$ powerlaw with a luminosity of $(1.4 \pm 0.3)
\times 10^{42} \ergps$. The nuclear spectrum is equally well fit with
a thermal model with a best fitting temperature suspiciously close to
that of the surrounding gas ($kT \sim 3\keV$).

Although noted in a few surveys, the radio source at the centre of the
Abell~2204 cluster, TXS1630+056 = 1632+0534, has received little
attention.  This is understandable given its relatively modest flux
and small angular size compared to most radio galaxies in cooling core
clusters.

At 8.4 GHz J1632+0534 consists of a compact nucleus with flux density
16.7 mJy with weaker compact components to the north and south.  The
total flux density is $\sim$22 mJy.  The nucleus has a flat spectrum.
Components on either side of the nucleus are steep spectrum, with some
evidence for extended, steep spectrum emission at 1.4 GHz with a
largest angular size of 15 arcsec (39 kpc).  A steep-spectrum
component, with flux density $\sim$1 mJy is present about 10 arcsec N
of the nucleus.  This component is likely associated with the nearby
companion galaxy seen in Fig.~\ref{fig:comparebands}.  At 1.4 GHz
J1632+0534 also extends to the south-west, and there is a hint of
diffuse emission 10 arcsec to the east and west that reaches 5 times
the rms noise level of 30 microJy/beam and could possibly be
associated with a mini-halo (Fig.~\ref{fig:comparebands}).  The total
flux density of J1632+0534 at 1.4 GHz is 58 mJy, and the corresponding
total radio power at 1.4 GHz is $3.7\times 10^{31} \erg \s^{-1}
\Hz^{-1}$.  Our 0.33 GHz (90~cm) image, with noise level $1
\:\textrm{mJy beam}^{-1}$, does not reveal any further extension, any
old emission from a previous outburst, or any sign of a mini-halo or
cluster halo.

The X-ray and radio nuclei are coincident with the optical nucleus of
one of the central galaxies. The two central galaxies are likely to be
associated because of the small velocity difference along the line of
sight \citep{Jenner74}.

The plateau contains a number of X-ray surface brightness
depressions. These include ones to the east, west, south-east and
north-east. There are also possible depressions to the south and
north. Each of the depressions has a radius of around 1.5~arcsec
(3.9~kpc), although the western hole could be larger with a radius of
around 2~arcsec (5.2~kpc). There is also the depression at the X-ray
centroid of the core and plateau, with a radius of around
1.5~arcsec. This depression lies around 2~arcsec south-east of the
nucleus.

Such depressions are seen in the cores of galaxy clusters, where the
relativistic gas in radio lobes displaces the thermal X-ray plasma
(e.g.  Hydra A, \citealt{McNamara00}; Perseus,
\citealt{BohringerPer93}, \citealt{FabianPer00}; Abell 2052,
\citealt{Blanton01}; Centaurus, \citealt{SandersCent02}). None of these
depressions are coincident with sources in our new deep radio data,
suggesting that if they are coincident with radio bubbles, the
electrons in the bubbles have aged too much to be observable in radio
(i.e. they are ``ghost'' cavities).

\subsection{Outer structure}
\begin{figure*}
  \includegraphics[width=0.7\textwidth]{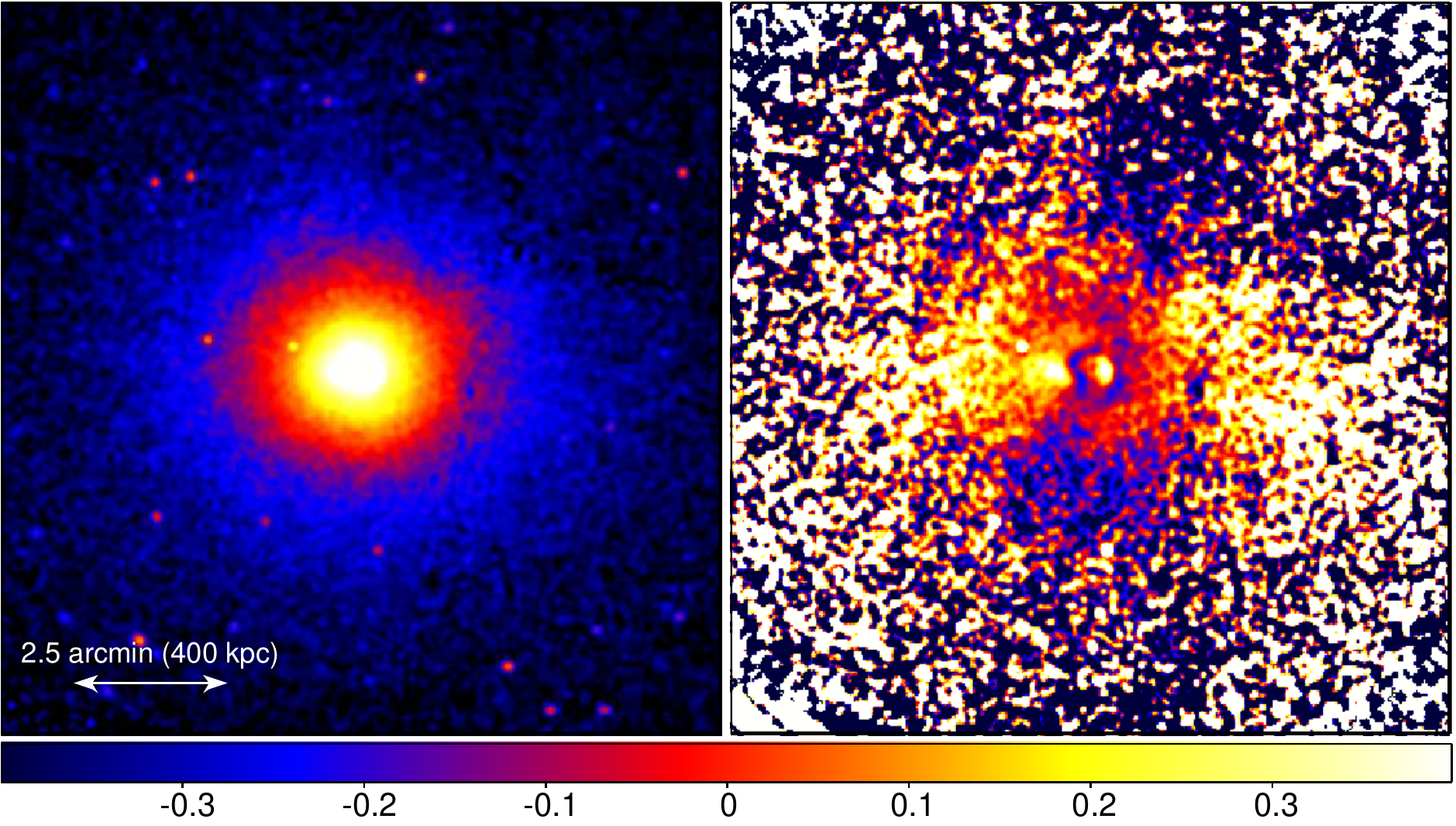}
  \caption{(Left) X-ray image of the larger scale cluster emission
    between 0.5 and 5~keV from the ACIS-I observations, binned into
    pixels of 2.0~arcsec and smoothed with a Gaussian with
    $\sigma=6$~arcsec. (Right) Fractional difference of each pixel
    from the average at that radius. The image was then smoothed with
    a $\sigma=6$~arcsec Gaussian. The colour bar shows the numerical
    scale of the fractional differences.}
  \label{fig:subavlarge}
\end{figure*}

We show in Fig.~\ref{fig:subavlarge} (left panel) the larger scale
surface brightness in the 0.5 to 5~keV band (generated from the lower
background ACIS-I observations). In this image we used the
\textsc{make\_readout\_bg} script (written by M. Markevitch) to
generate out-of-time events background images to subtract from the
data. Out-of-time events occur while the detector is being read,
leading to a streak in the readout direction (the exposure of these
events is effectively 1/78.05 of the normal exposure). Without
subtraction, the streak runs about 12 degrees north from the west
through the centre of the cluster for the 7940 observation
(coincidentally running along the plume).

In Fig.~\ref{fig:subavlarge} (right panel) we show the fractional
deviation of each pixel in the surface brightness image from the
average at that radius (accounting for the effects of background and
out-of-time events). Immediately apparent are enhancements in roughly
the east-west direction, and depressions in the north-south
direction. 

\begin{figure}
  \centering
  \includegraphics[width=\columnwidth]{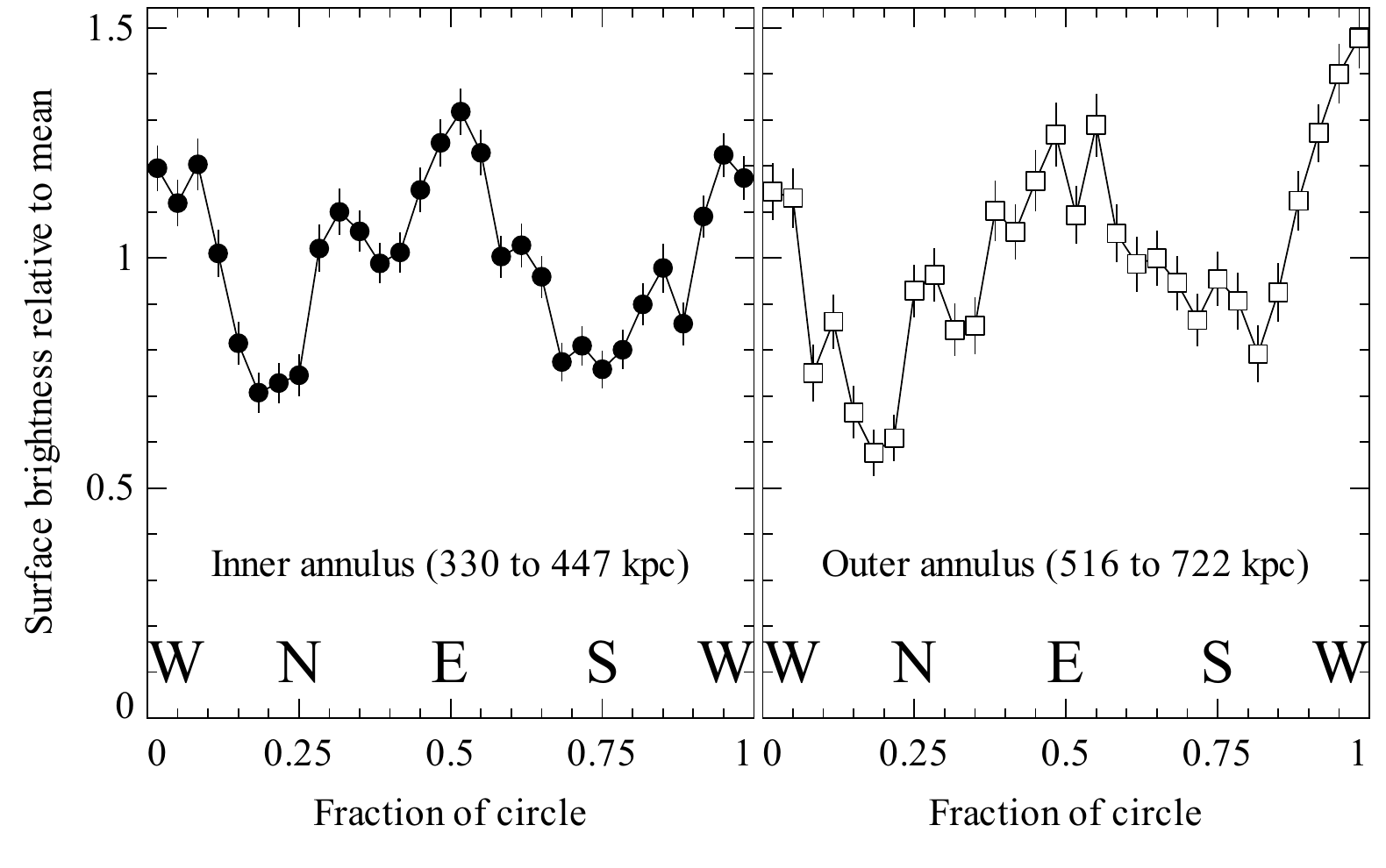}
  \caption{Surface brightness profile around in azimuth at two radii,
    chosen to overlap with the inner decrement to the south and outer
    decrement to the north. The profiles have been normalised by the
    average at each radius.}
  \label{fig:aziprofiles}
\end{figure}

\begin{figure}
  \centering
  \includegraphics[width=0.8\columnwidth]{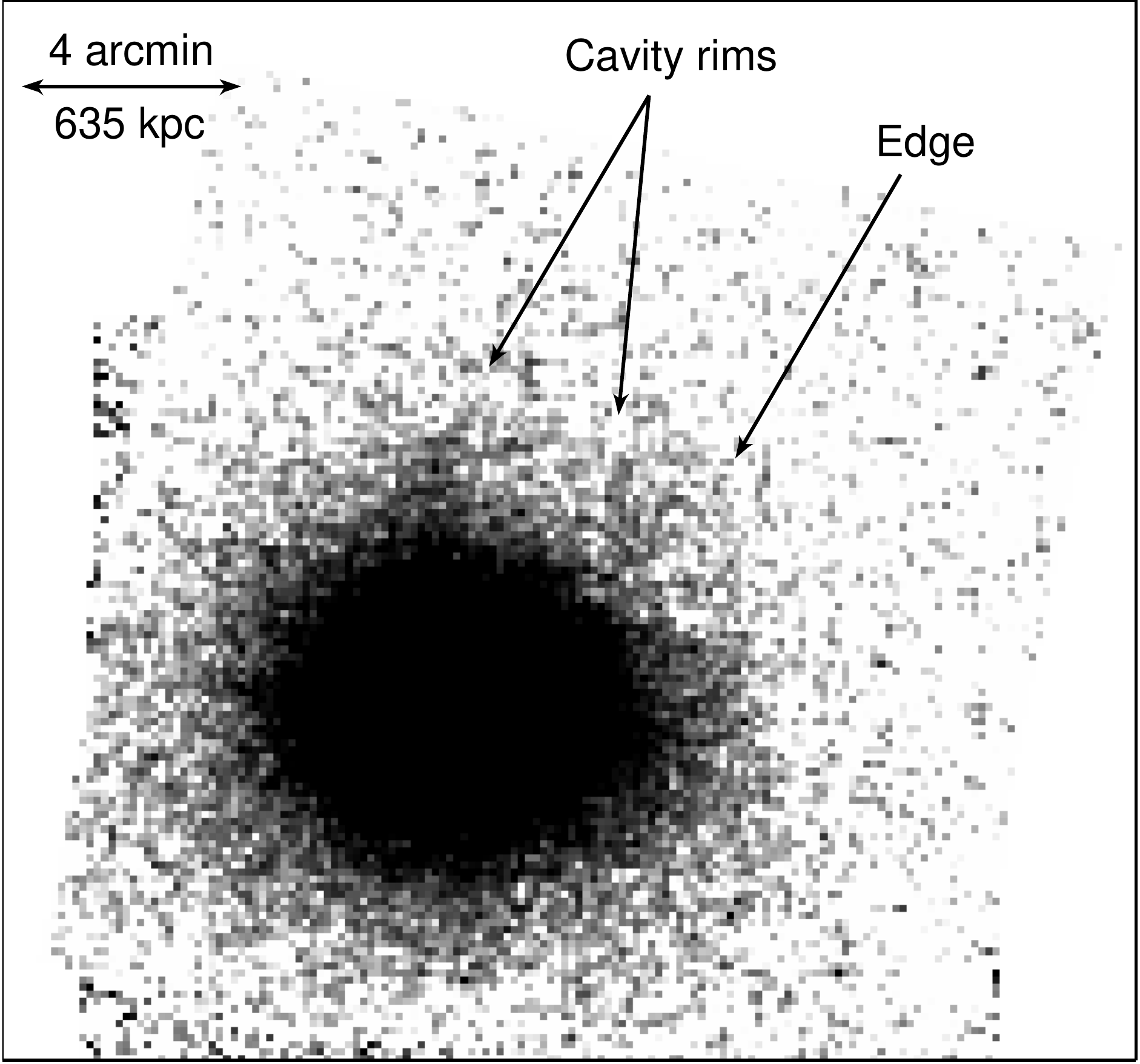}
  \caption{0.5 to 5.0 keV exposure-corrected and background-subtracted
    image of the outer region of the cluster, with point sources
    excluded, binned using 7.87 arcsec pixels and smoothed with a
    Gaussian of 1 pixel.}
  \label{fig:outerimg}
\end{figure}

To examine these in more detail, Fig.~\ref{fig:aziprofiles} displays
an azimuthal surface brightness profiles at two different radii,
chosen to coincide with the inner (southern) and outer (northern)
decrements. The decrements are very significant, at around the 30
per~cent level.

The northern decrement and its surrounding bright rims can easily be
seen in an image of the cluster with a large binning factor applied
(Fig.~\ref{fig:outerimg}). In addition, an edge like feature is
apparent 950~kpc to the north-west of the cluster, close to the
maximum extent of the northern decrement. This edge is also just
visible in Fig.~\ref{fig:subavlarge}. \emph{XMM-Newton} MOS images of
the cluster also show the northern bubble, though its edge lies close
to the edge of the central MOS detector.

The most important question is the physical nature of these decrements
in Fig.~\ref{fig:subavlarge} and Fig.~\ref{fig:aziprofiles}. The first
option is that they are an artefact of the analysis procedure. If the
cluster has an elliptical surface brightness distribution on the sky,
then if we subtract the average surface brightness at each radius we
will obtain a sinusoidal profile. Numerical tests show that we need
approximately a 2:1 ratio of major to minor axis to achieve the 30
per~cent surface brightness variation observed. Fitting elliptical
surface brightness models to the data imply that the cluster is fairly
close to circular (see also \citealt{Hashimoto07} who find a low
ellipticity), excluding the very inner regions. The profiles
(Fig.~\ref{fig:aziprofiles}) are also not very sinusoidal,
particularly the sharp edges to the depression to the
north. Fig.~\ref{fig:outerimg} shows that there are real surface
brightness enhancements surrounding the northern decrement, indicating
it is not a data analysis issue.

Therefore the depressions are likely to be real cluster
features. Their morphology is similar to the X-ray cavities observed
in other galaxy clusters. In this case the cavities are very large
compared to those typically found: the northern cavity has a radius of
around 85~arcsec (225~kpc) and the southern one 55~arcsec
(145~kpc). If they are cavities, then they are significantly larger
than the 100~kpc radius cavities in MS~0735+7421 \citep{McNamara05}
and Hydra~A \citep{Wise07}. The cavity interpretation is favoured by
the enhancement in surface brightness around the depression,
particularly for the northern feature (see
Fig.~\ref{fig:outerimg}). These enhancements would correspond to the
bright rims seen around bubbles in other objects.

Instead of a single episode, the depressions may be the cumulative
result of a set of repeated outbursts along a single axis.  These
outbursts could accumulate into a large region occupied by non-thermal
particles, displacing the thermal plasma. Repeated sets of bubbles
along one direction have been seen in Hydra~A \citep{Wise07},
estimated to have lasted over 200--500 Myr. In addition the Perseus
cluster shows a low-thermal-pressure channel and further possible
bubbles to the south beyond the outermost southern radio bubble. There
is also a possible intact bubble 170~kpc to the north
\citep{SandersPer07}.

Bubbles should ``pancake'' at large radius \citep{Churazov01}, where
the bubble density matches the ambient density. It may be that they
have not yet done so here, perhaps because of magnetic fields.

Other possibilities include a merger. However, the cavities with their
sharp edges appear unlikely to be the result of a merger.  Finally, we
could perhaps be observing the accretion of matter onto the cluster
along filaments to the east and west of the image. This could be a
possibility as the temperature of the gas is enhanced along these
directions (see Section \ref{sect:outerspectra}). However the
temperature enhancement can only be measured close to the centre of
the cluster, not where the material would be being accreted. Also the
sharp edges of the depressions make this explanation unlikely.

\section{Spectral fitting}
\subsection{Mapping the cluster core}
\label{sect:mapcore}
To investigate the metallicity and temperature structure in the core
of the cluster, we applied spatially resolved spectroscopy
techniques. We divided the central $\sim 70 \times 70$~arcsec into
bins using the Contour Binning algorithm \citep{SandersBin06} which
follows surface brightness variations. We chose regions with a
signal-to-noise ratio of 30 (around 900 counts), restricting the
length of the bins to be at most two times their width.

We extracted spectra from each of the datasets, generating appropriate
responses and ancillary responses. Background spectra were extracted
from the part-synthetic backgrounds (Section \ref{sect:bg}). The 6104
and 7940 datasets were added together (as they use the same detector),
weighting responses and backgrounds.

We fit each spectrum with an absorbed \textsc{apec} model
\citep{SmithApec01}. The absorption was fixed to the Galactic value,
but the temperature, metallicity and emission-measure were allowed to
vary in the fit. The fitting procedure minimised the C-statistic
\citep{Cash79} and we fitted the data between 0.5 and 7~keV.

\begin{figure*}
  \includegraphics[width=0.30\textwidth]{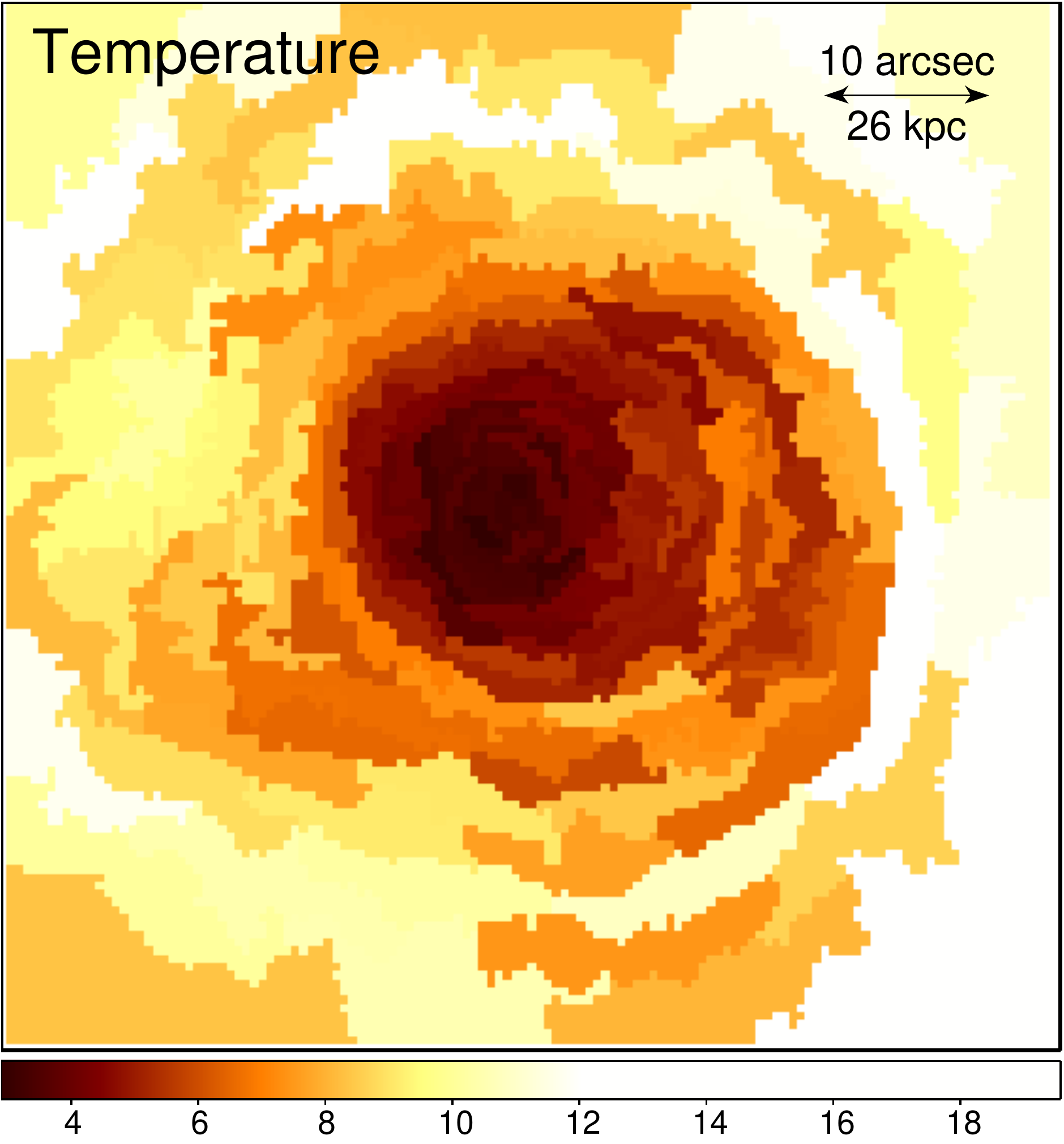}
  \includegraphics[width=0.30\textwidth]{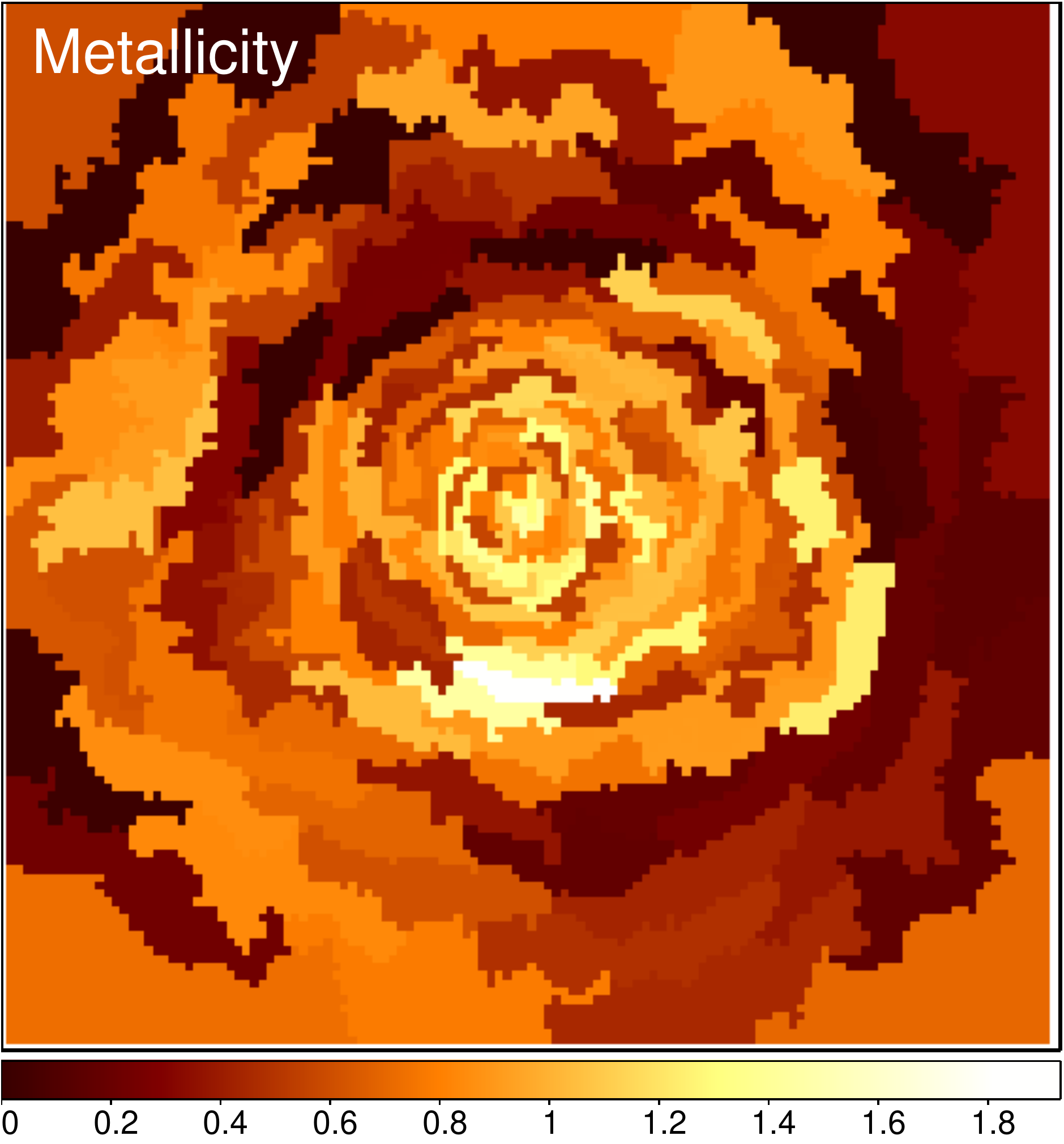}
  \includegraphics[width=0.30\textwidth]{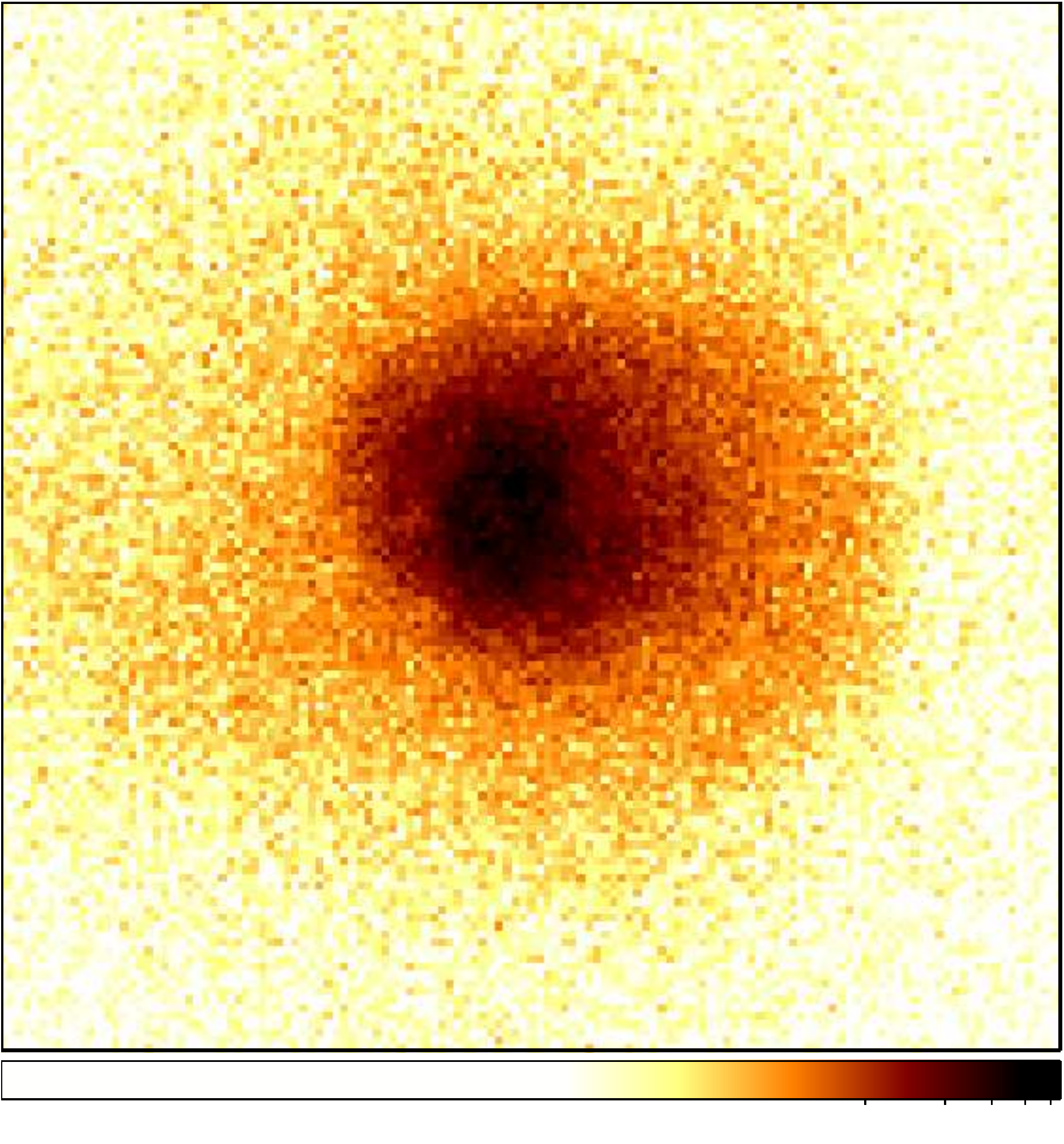}
  \caption{Temperature (left), metallicity (centre) and image (right)
    of the core of the cluster at the same spatial scale. The units of
    temperature are KeV and metallicity $\Zsun$. These maps were
    created by fitting models to spectra with around 900 counts. 10
    arcsec corresponds to a distance of $\sim 26$~kpc at this
    redshift. Uncertainties in temperature range from around 6
    per~cent in the centre to 30 per~cent in the outer parts of this
    image. Metallicities are uncertain to around 0.35\Zsun.}
  \label{fig:zoommaps}
\end{figure*}

Shown in Fig.~\ref{fig:zoommaps} are the derived temperature and
metallicity maps of the core of the cluster, with an image of the
cluster at the same scale. Note that the temperature is
emission-weighted and projected and that the metallicity measurements
are most sensitive to iron. The bright core and plateau corresponds to
a cool region, where the projected temperature drops below 3~keV in
some places. There also appears to be an extension in cool gas along
the plume to the west of the core.

\begin{figure}
  \centering
  \includegraphics[width=0.9\columnwidth]{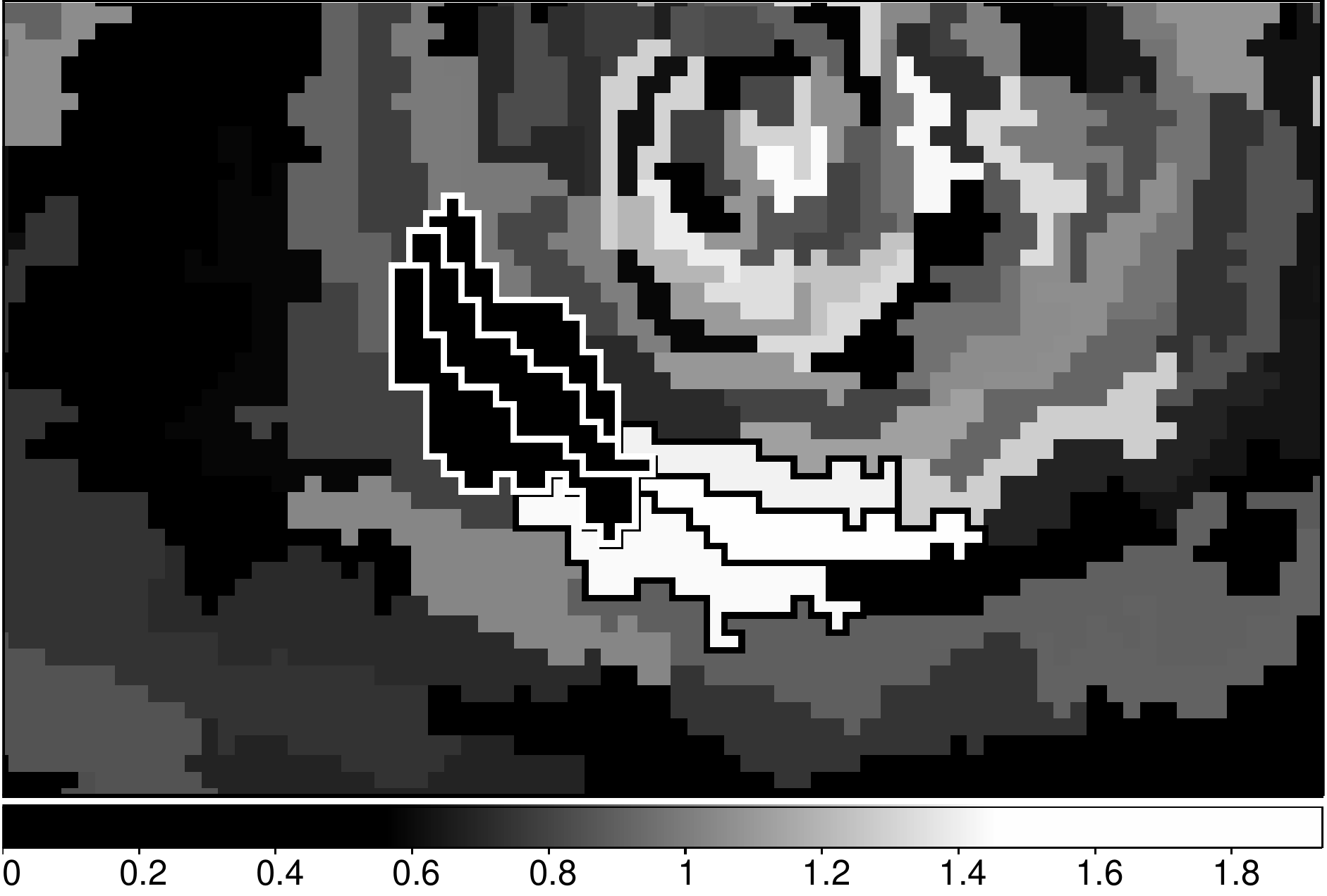}
  \caption{Zoom up of the metallicity map showing the three adjacent
    high metallicity regions (bordered in black) and the three
    adjacent low metallicity regions (bordered in white), examined in
    Section~\ref{sect:mapcore} and Fig.~\ref{fig:highzcentre}. The
    units are $\Zsun$ and each pixel is 0.49~arcsec (1.3~kpc) in
    size.}
  \label{fig:zoommetals}
\end{figure}

\begin{figure}
  \includegraphics[width=\columnwidth]{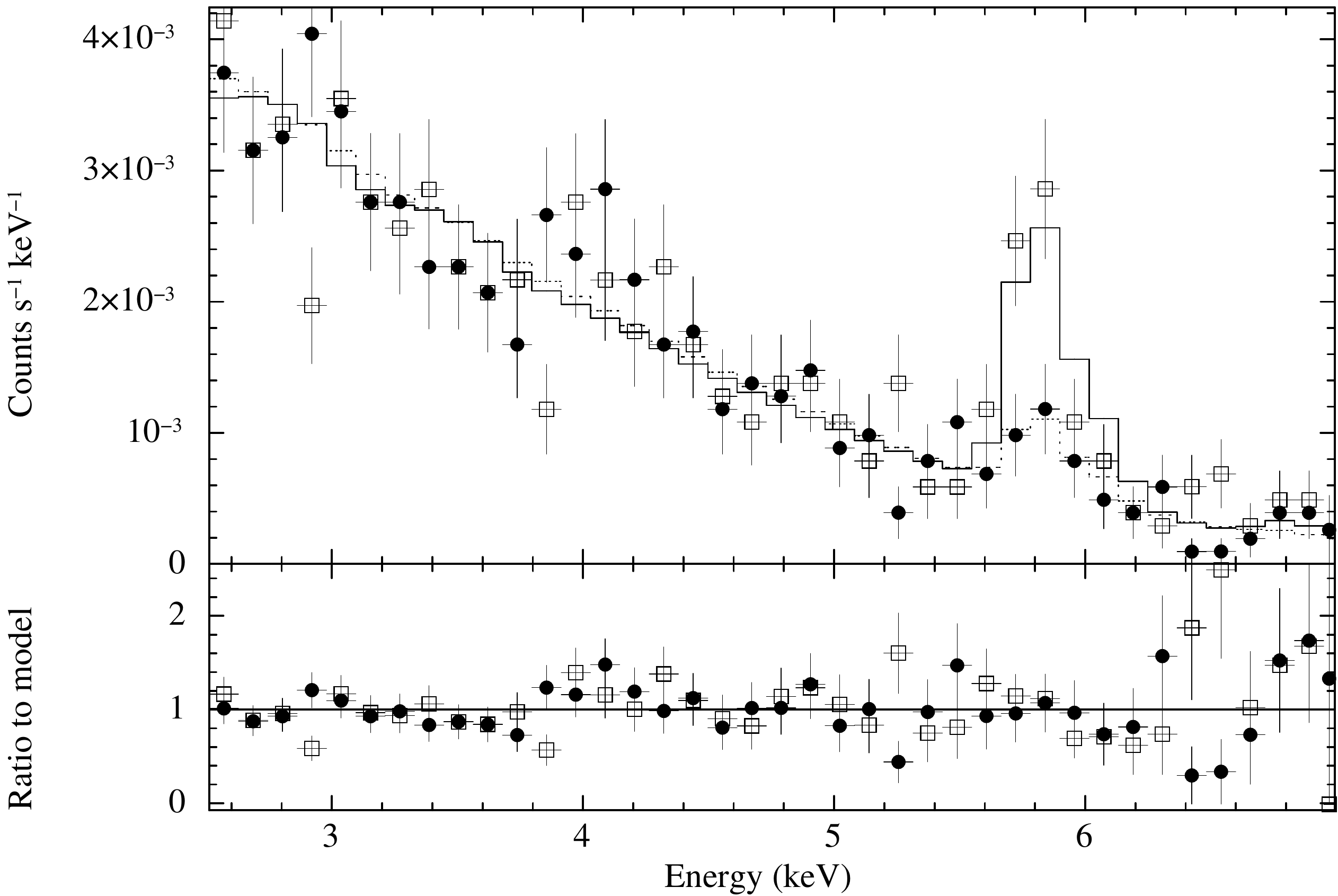}
  \caption{Comparison of the rebinned higher energy part of the
    spectrum between the high and low metallicity regions 11~arcsec to
    the south and south-east of the core
    (Fig.~\ref{fig:zoommaps}). The solid and dotted lines are the best
    fitting models. The bottom panel shows the ratio of the rebinned
    data to model.}
  \label{fig:highzcentre}
\end{figure}

The metallicity structure appears to have a different morphology. The
core and plateau contains a number of metallicity inhomogeneities. The
most significant is a high metallicity region around 11~arcsec to the
south of the centre (Fig.~\ref{fig:zoommetals}). This is adjacent to a
region with low metallicity to the south-east of the core at the same
radius. We show the combined spectrum in the three high metallicity
bins compared to the three low metallicity bins to their east in
Fig.~\ref{fig:highzcentre}. The metallicity of the low metallicity
region is $0.38^{+0.17}_{-0.14} \Zsun$ and the high metallicity region
is $1.60^{+0.39}_{-0.32} \Zsun$. If both regions are constrained to
have the same temperature, this makes very little difference to this
result. Markov Chain Monte Carlo (MCMC) tests in \textsc{xspec}
confirm these uncertainties.

Looking at the spectra (Fig.~\ref{fig:highzcentre}), the significance
of the enhancement in the three high metallicity bins relative to the
three low metallicity bins is around $4\sigma$ (corresponding to a
probability of $6 \times 10^{-5}$ of occurring by chance). There are
168 bins in the image, so there are approximately 170 sets of three
radial bins and around 340 unique sets of three radial bins next to
three radial bins. This increases the chance probability to around 2
percent.

However, the highest, second highest and fifth highest metallicity
bins are next to each other. They are statistically independent bins,
so the chances of this occurring by chance are the order of $(2/167) (3
\times 2/164) \sim 4 \times 10^{-4}$. This small probability is
decreased still further by the chances of finding three of the lowest
metallicity regions in the central part of the cluster immediately
adjacent to these high metallicity regions.

\subsection{Mapping the outer regions of cluster}
\label{sect:outerspectra}
\begin{figure}
  \centering
  \includegraphics[width=0.8\columnwidth]{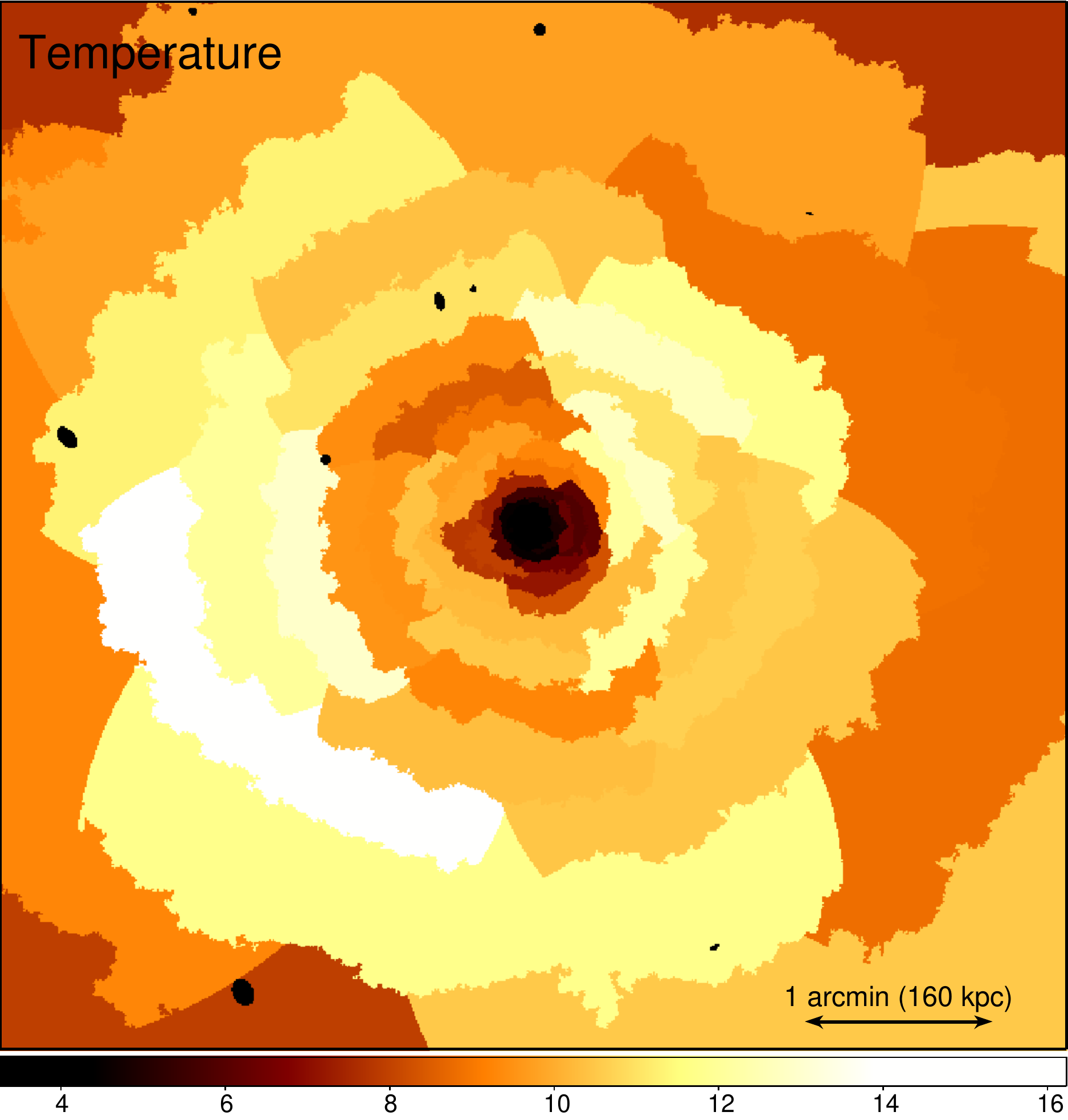}
  \vspace{4mm}
  \includegraphics[width=0.8\columnwidth]{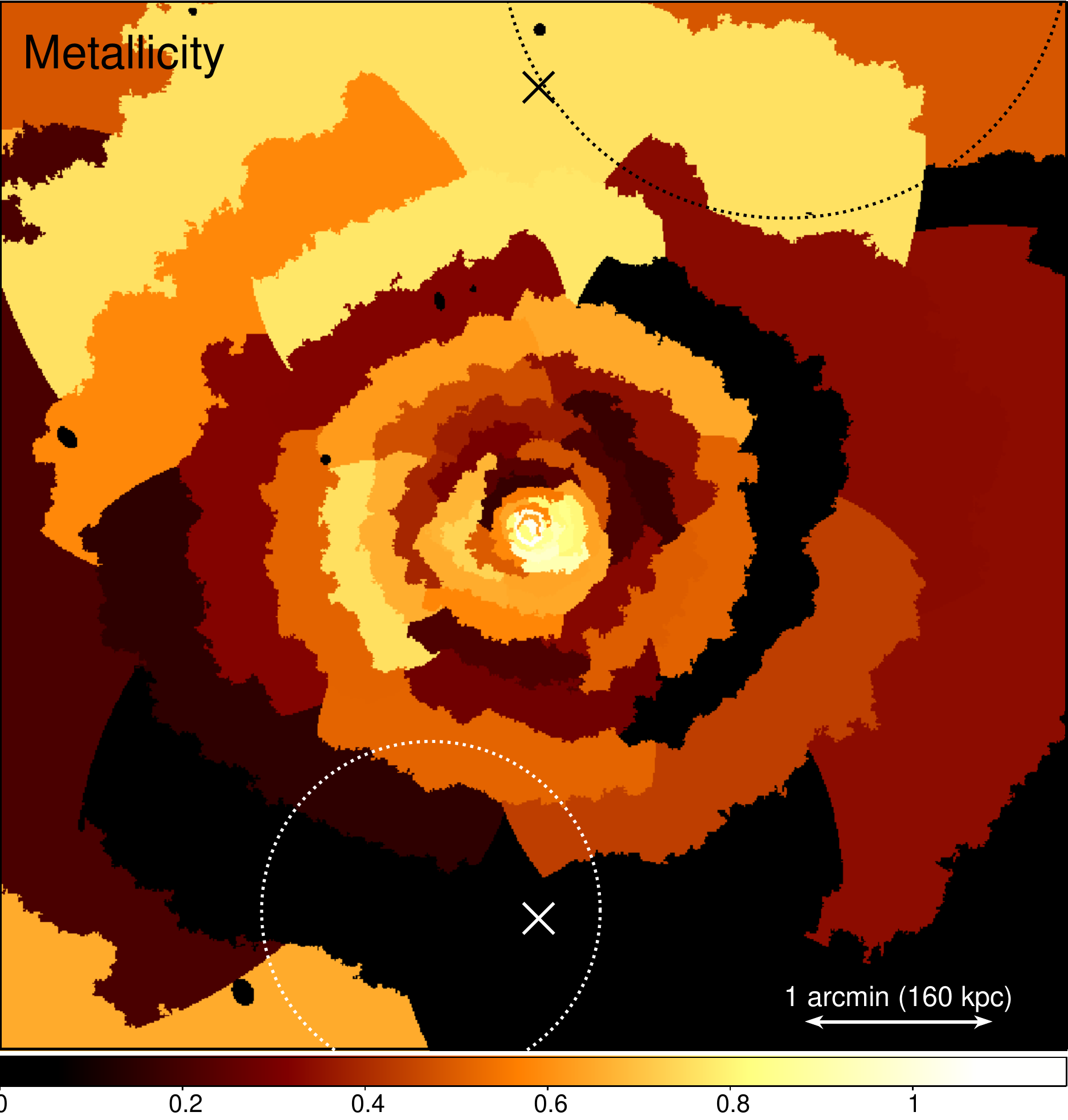}
  \caption{Larger scale temperature (top; in keV) and metallicity
    (bottom; in Solar units) structure, created by fitting spectra
    with a signal-to-noise ratio of greater than 63. The two regions
    marked with an ``X'' are examined spectrally in
    Fig.~\ref{fig:outerspeccompar}. The outer surface brightness
    depressions are marked as dotted circles on the metallicity map.}
  \label{fig:outermaps}
\end{figure}

We show in Fig.~\ref{fig:outermaps} temperature and metallicity maps
of a $\sim 870 \times 870$~kpc box around the cluster core. To
generate these maps, we binned the image to have regions with a
signal-to-noise ratio of 63 ($\sim 4000$ counts) with the Contour
Binning algorithm. We generated combined spectra from the 6104 and
7940 ACIS-I observations, fitting the data with the same \textsc{apec}
model as for the core region.

We investigated the effect of accounting for out-of-time events
(events occurring while the CCDs are being read out), but this made
little difference to the spectral fitting results for this
region. Minimising $\chi^2$ instead of the C-statistic made some
difference, but the morphology remained very similar.

The temperature map shows that the temperature of the gas to the east
flattens off to around 8~keV, until a radius of around 190~kpc, where
it steeply rises to above 12~keV. The temperature rises more steeply
to the west. At large radius there is an apparent decline in
temperature.

The metallicity map shows structure, as we found previously
\citep{SandersA220405}. We find that the metallicity drops outside of
the core, but rises again to form a ring or spiral around the core at
a radius of around 150~kpc. It is unclear whether this metallicity
structure is connected to the spiral-like feature in the very central
regions (Fig.~\ref{fig:zoommaps} centre panel). 

\begin{figure}
  \includegraphics[width=\columnwidth]{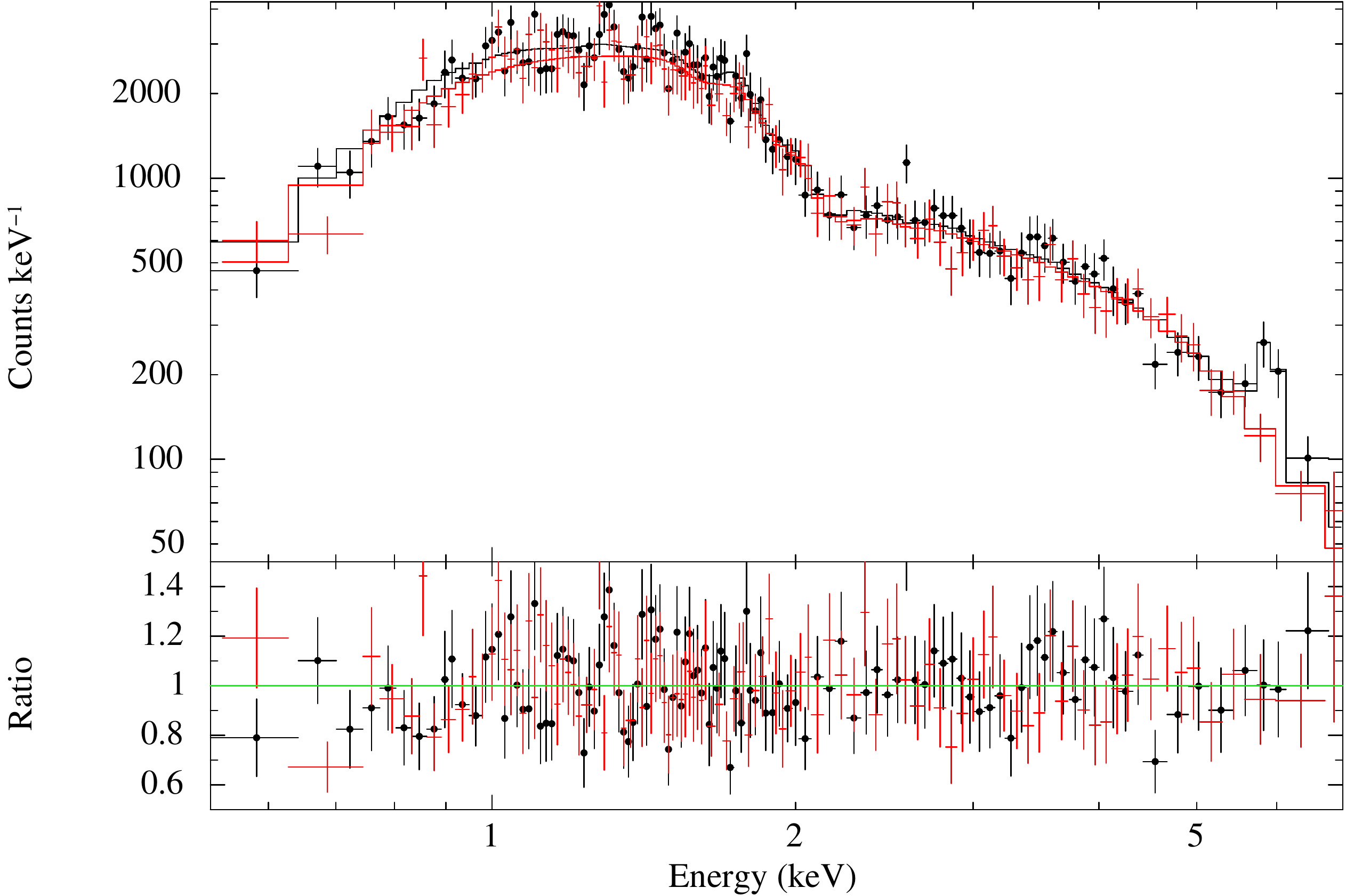}
  \caption{Comparison of the spectra from regions around 350~kpc to
    the north (circles) and south (no circles). The solid lines in the
    top panel are the best fitting spectral models. In the bottom
    panel is shown the ratio of the data to the best fitting spectral
    model. The data have been rebinned.}
  \label{fig:outerspeccompar}
\end{figure}

One particularly interesting metallicity feature is the region of high
metallicity around 300~kpc to the north. There is no high metallicity
material at this radius to the south. This can be demonstrated by
examining the spectra directly of two diametrically opposite bins
directly (Fig.~\ref{fig:outerspeccompar}; the bins examined are marked
in Fig.~\ref{fig:outermaps}). There is no evidence for Fe-K line
emission for the southern spectrum, but it is seen strongly for the
northern region.

The best fitting metallicities are zero for the southern region (the
$2\sigma$ upper limit is $0.28 \Zsun$). The metallicity of the
northern region is $0.77\pm 0.21$. We have confirmed with the MCMC
functionality in \textsc{xspec} that the difference in metallicity
between the two regions is significant to $3\sigma$.

\subsection{Profiles of cluster properties}
To examine the cluster properties as a function of radius, we
extracted projected spectra from circular annuli chosen to give a
reasonable quality spectrum. We only used the 7940 dataset here, as it
simplified much of the analysis without much change in signal-to-noise
ratio. Using the Direct Spectral Deprojection method of
\cite{SandersPer07} (tested in \citealt{Russell08}), we generated
deprojected spectra from the projected spectra. We then fitted single
temperature \textsc{apec} models to the projected and deprojected
spectra to create temperature, metallicity and emission measure
profiles. We included the effects of background and out-of-time events
in this analysis, assumed Galactic absorption, grouped the spectra to
have at least 25 counts per spectral bin, fitted the data between 0.5
and 7~keV and minimised the $\chi^2$ to find the best fitting model.

\begin{figure}
  \includegraphics[width=\columnwidth]{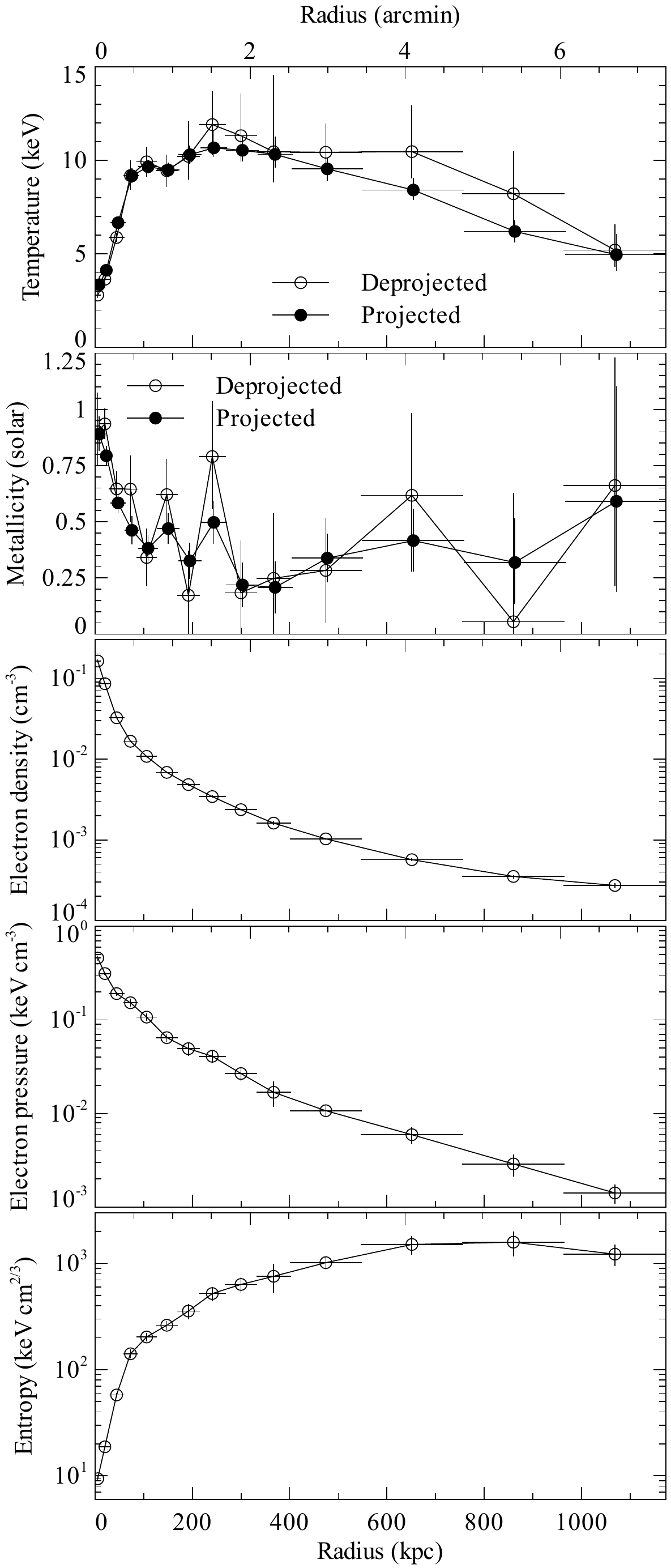}
  \caption{Temperature, metallicity and electron density profiles for
    the outer regions of the cluster. Also shown are the computed
    electron pressure and entropy profiles.}
  \label{fig:profiles}
\end{figure}

Fig.~\ref{fig:profiles} shows the projected and deprojected
temperature and projected metallicity profiles out to a radius of
$\sim 1200$~kpc. Beyond this radius there is little cluster emission
in the spectra. The plot also shows the electron density, calculated
from the emission measure of the fit to the deprojected spectra. By
multiplying the deprojected electron density and temperature, we
calculated the pressure. We also computed the entropy using $S = kT
n_\mathrm{e}^{-2/3}$.  To calculate the errors on the pressure and
entropy, we assumed that the errors on temperature and density were
independent. Most of the uncertainty is in the temperature, so this
assumption could be made.

We are able to measure a density variation of a factor of $\sim 600$,
pressure $\sim 300$ and entropy $\sim 150$. We observe the temperature
of the ICM to decline with radius beyond 400~kpc (in the projected
spectra, which have smaller error bars). The temperature profile in
the outskirts agrees reasonably well with the \emph{Suzaku},
\emph{Chandra} and \emph{XMM-Newton} profiles published by
\cite{Reiprich08}. In the centre, we find values similar to their
\emph{Chandra} results from a shorter observation. The peak
\emph{Chandra} temperatures are higher than \emph{Suzaku} or
\emph{XMM} and the central temperature drops more steeply. This is
probably due to PSF effects, though there are claims that the
\emph{Chandra} effective area calibrations contribute to the higher
peak temperatures. Interestingly, our results show that the entropy
profile remains flat beyond 550~kpc, similar to what was found in
PKS~0745-191 \citep{George08}. The entropy profile appears to break at
72, 240 and 650~kpc.

\begin{figure}
  \includegraphics[width=\columnwidth]{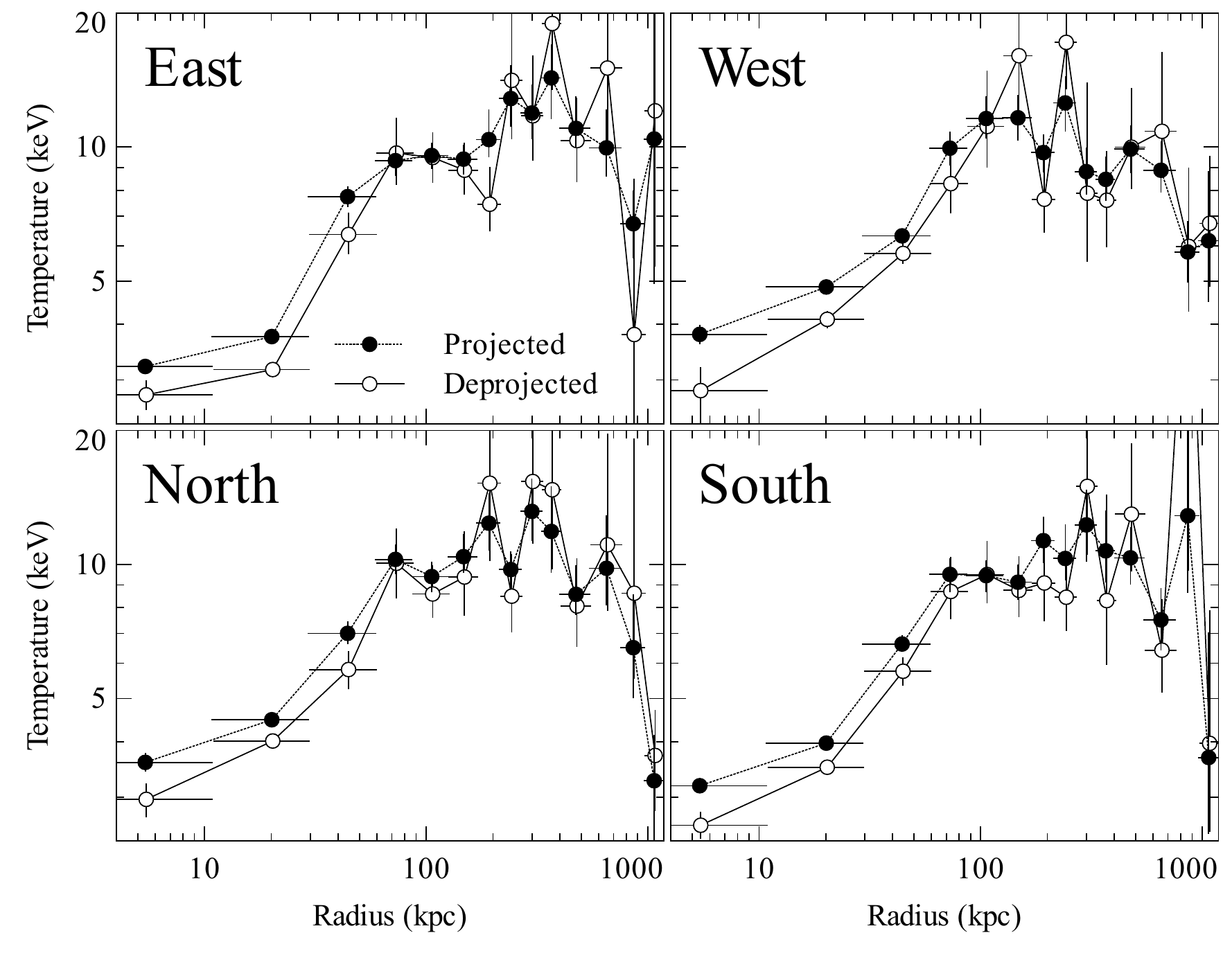}
  \caption{Projected and deprojected temperature profiles in four
    sectors.}
  \label{fig:quadT}
\end{figure}

\begin{figure}
  \includegraphics[width=\columnwidth]{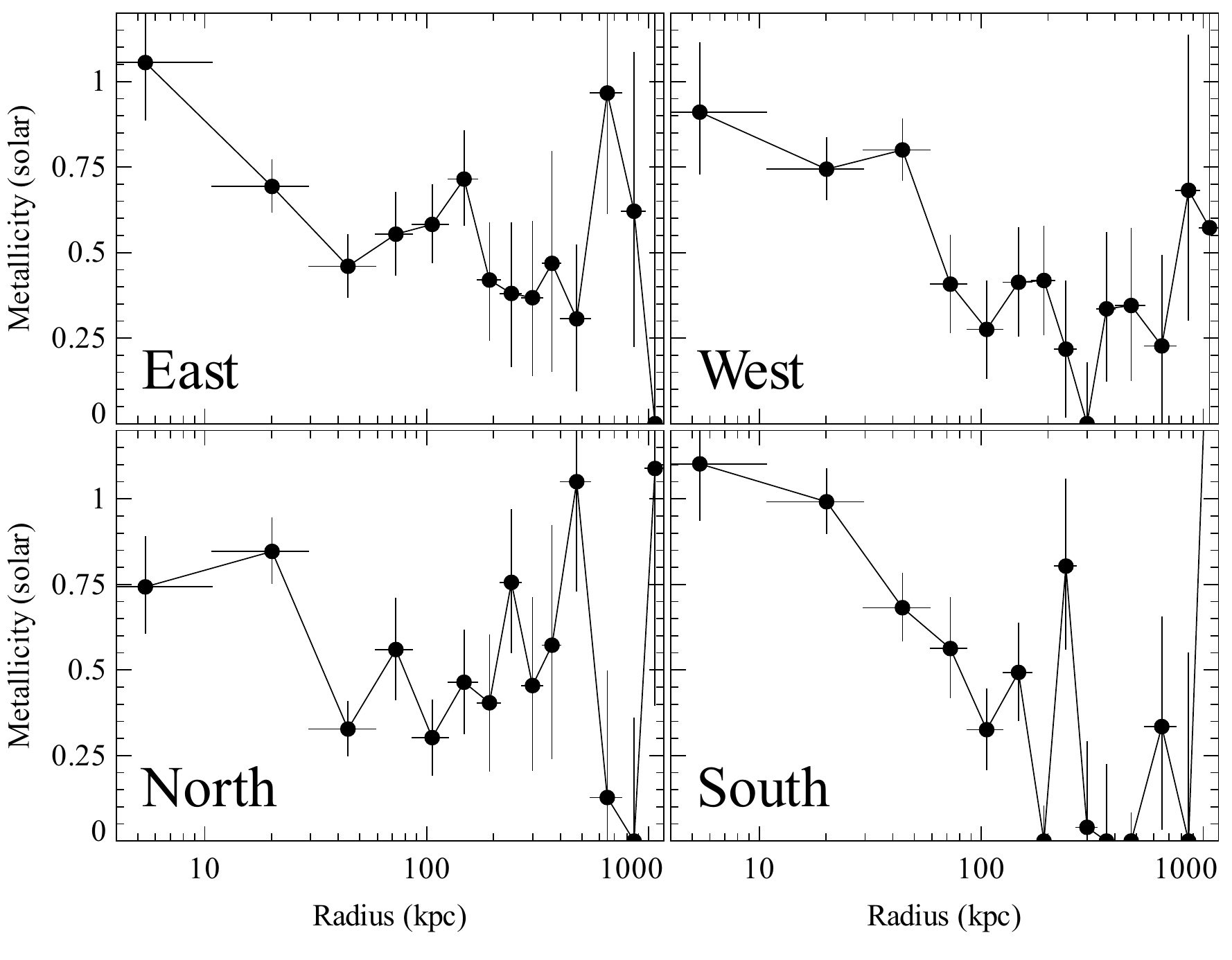}
  \caption{Projected metallicity profiles in four sectors.}
  \label{fig:quadZ}
\end{figure}

\begin{figure}
  \includegraphics[width=\columnwidth]{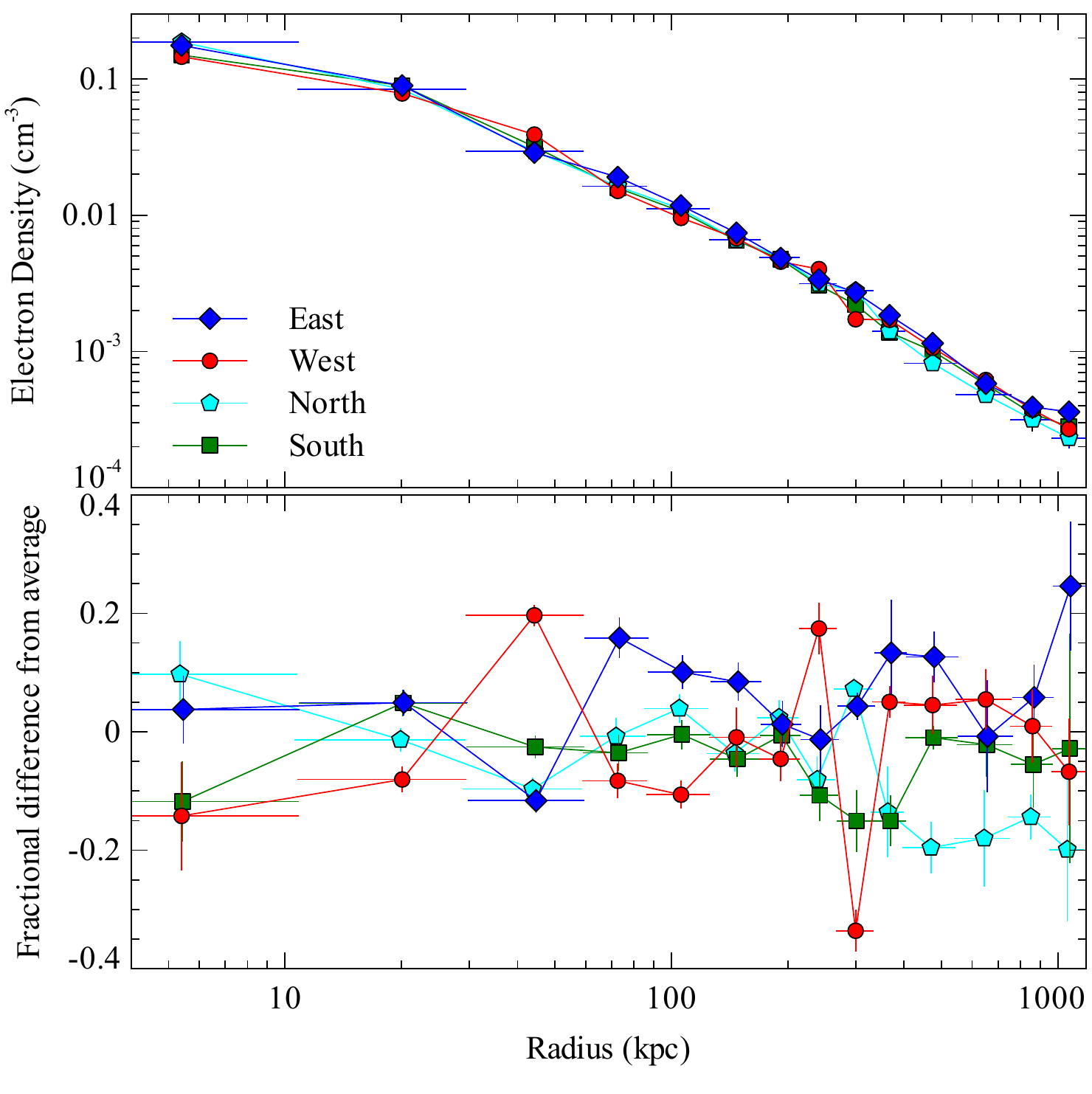}
  \caption{Deprojected electron density profiles in four sectors. The
    bottom panel shows the fractional difference of each profile to
    the weighted mean at each radius.}
  \label{fig:quadne}
\end{figure}

The maps suggest that there is structure in the properties of the
cluster as a function of angle. We split up each annulus into four 90
degree sectors, pointing towards the north, east, south and
west. Fig.~\ref{fig:quadT} shows the projected and deprojected
temperature profiles for each of these sectors. Much of the structure
in the temperature maps is seen in the profiles. For instance, 120~kpc
to the west is a higher temperature region. At larger radius to the
west, where there is brighter cluster emission than to the north or
south (Fig.~\ref{fig:outerimg}), the gas is cooler than in the other
sectors.

The projected metallicity profiles in the quadrants
(Fig.~\ref{fig:quadZ}) show interesting variation. We see the high
metallicity ring strongly to the south at a radius of around
240~kpc. We also see the large metallicities to the north 650~kpc from
the nucleus.

In Fig.~\ref{fig:quadne} (top panel) are plotted the deprojected
densities in the four sectors. In the bottom panel we display the
fractional difference of each density from the average at that
radius. The plot shows the lower density regions to the north (beyond
300 kpc) and south (around 250~kpc) corresponding to the depressions
in the surface brightness image.

\subsection{Cool gas in the core}
To examine the amount of cool gas which could be present in the core
of the cluster, we extracted the spectrum from the inner 100~arcsec
radius. We fit a cooling flow model, made up of \textsc{apec+mkcflow}
components to account for the cluster and cooling emission. The
\textsc{mkcflow} cooling flow model was computed with \textsc{apec}
spectra. Galactic absorption was assumed in the spectral fitting. The
upper temperature and metallicity of the cooling flow model was tied
to the \textsc{apec} component and the lower temperature was fixed to
0.0808~keV (the lowest possible). The \emph{Chandra} spectra were
consistent with a $65 \pm 21 \Msunpyr$ mass deposition rate.

\begin{table}
  \caption{\emph{XMM-Newton} RGS datasets analysed. Exposures given are
    for the RGS1 instrument.}
  \begin{tabular}{lll}
    Observation ID & Date & Exposure (ks) \\ \hline
    0112230301 & 2001-09-12 & 21.4 \\
    0306490101 & 2006-02-06 & 23.4 \\
    0306490201 & 2006-02-08 & 23.5 \\
    0306490301 & 2006-02-12 & 23.1 \\
    0306490401 & 2006-02-14 & 22.8 \\
  \end{tabular}
  \label{table:xmmdatasets}
\end{table}

Much better determinations of the amount of cool gas can be made with
high spectral resolution \emph{XMM-Newton} RGS data (see
\citealt{PetersonFabian06} and references therein). We processed each
of the RGS observations of Abell~2204 (Table~\ref{table:xmmdatasets})
with \textsc{sas} version 8.0.0. We used a PSF extraction region of 90
per cent and a pulse-height distribution region of 95 per cent.  We
created combined first-order RGS1 and combined RGS2 spectra and
responses using the \textsc{rgscombine} task. The data were grouped to
have at least 25 counts per spectral bin. Background model spectra
were created with \textsc{rgsbkgmodel}.

We fit the first-order data between 7 and 26~{\AA} with a
multi-temperature model. The temperature components were
\textsc{vapec} models with fixed temperatures of 0.25, 0.5, 1, 2, 4
and 8~keV but free normalizations. The components shared the same
metallicities, with O, Ne, Mg, Si, Fe and Ni free in the fits, and
with S, Ar and Ca tied to Fe. The components were absorbed with fixed
Galactic absorption. We did not account for the spatial distribution
of the source in the modelling as the bright region is small compared
to the \emph{XMM} PSF. We fit the model to minimize the $\chi^2$
statistic. The reduced $\chi^2$ of the best fitting model was $0.994 =
1273.56/1281$.

\begin{figure}
  \centering
  \includegraphics[width=\columnwidth]{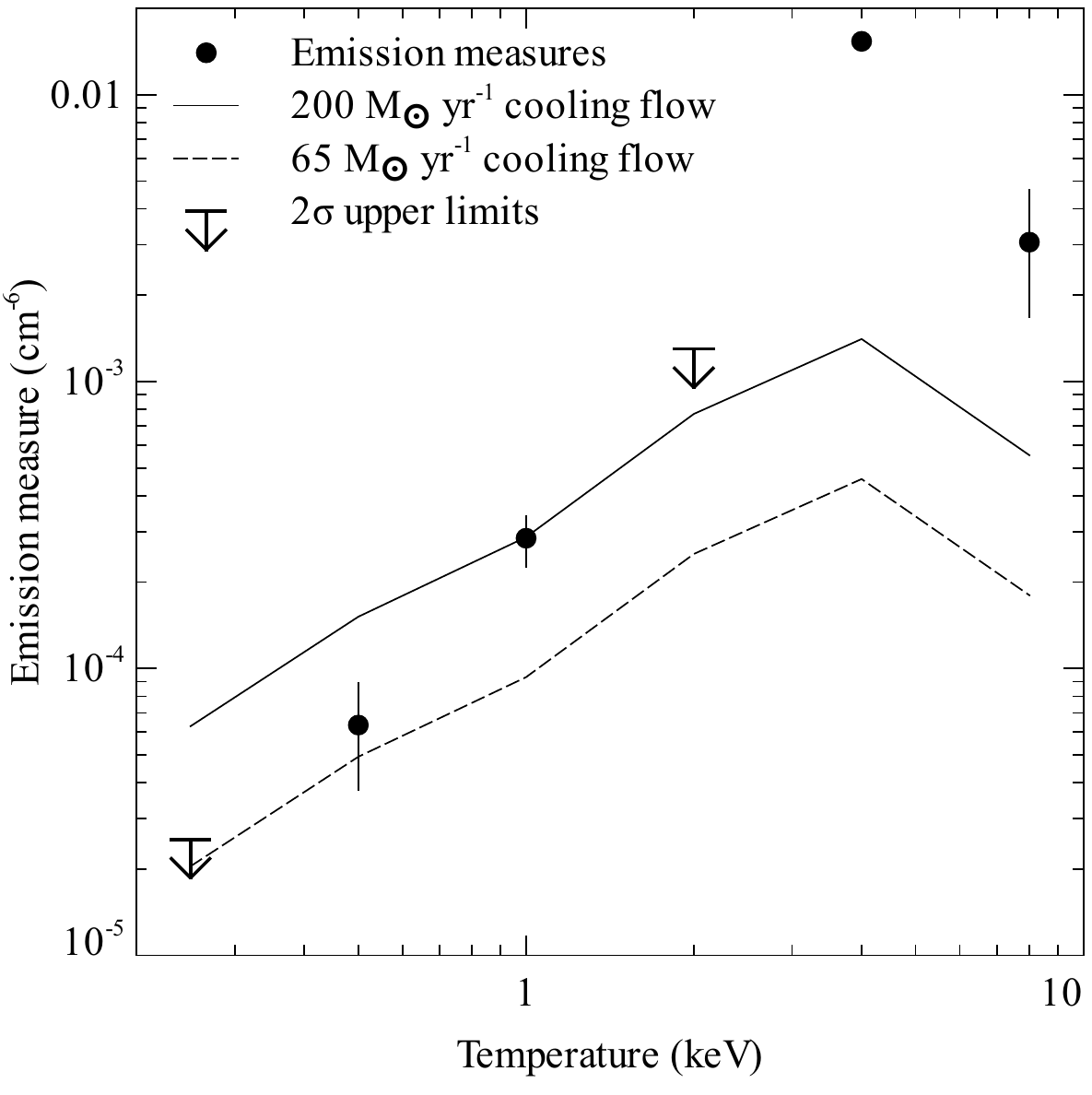}
  \caption{Best fitting emission measures for each of the components
    in the spectral fit to the RGS data.}
  \label{fig:rgsnorms}
\end{figure}

In Fig.~\ref{fig:rgsnorms} we show the best fitting emission measures
for each of the temperature components (these are the normalizations
produced by \textsc{xspec}). Also plotted are the emission measures
found by fitting a simulated spectrum of a cooling flow with a mass
deposition rate of $200 \Msunpyr$ cooling from 8 to 0.0808~keV. We
also plot the normalizations for a $65 \Msunpyr$ cooling flow, as
found from the \emph{Chandra} spectra.

\begin{figure}
  \centering
  \includegraphics[width=\columnwidth]{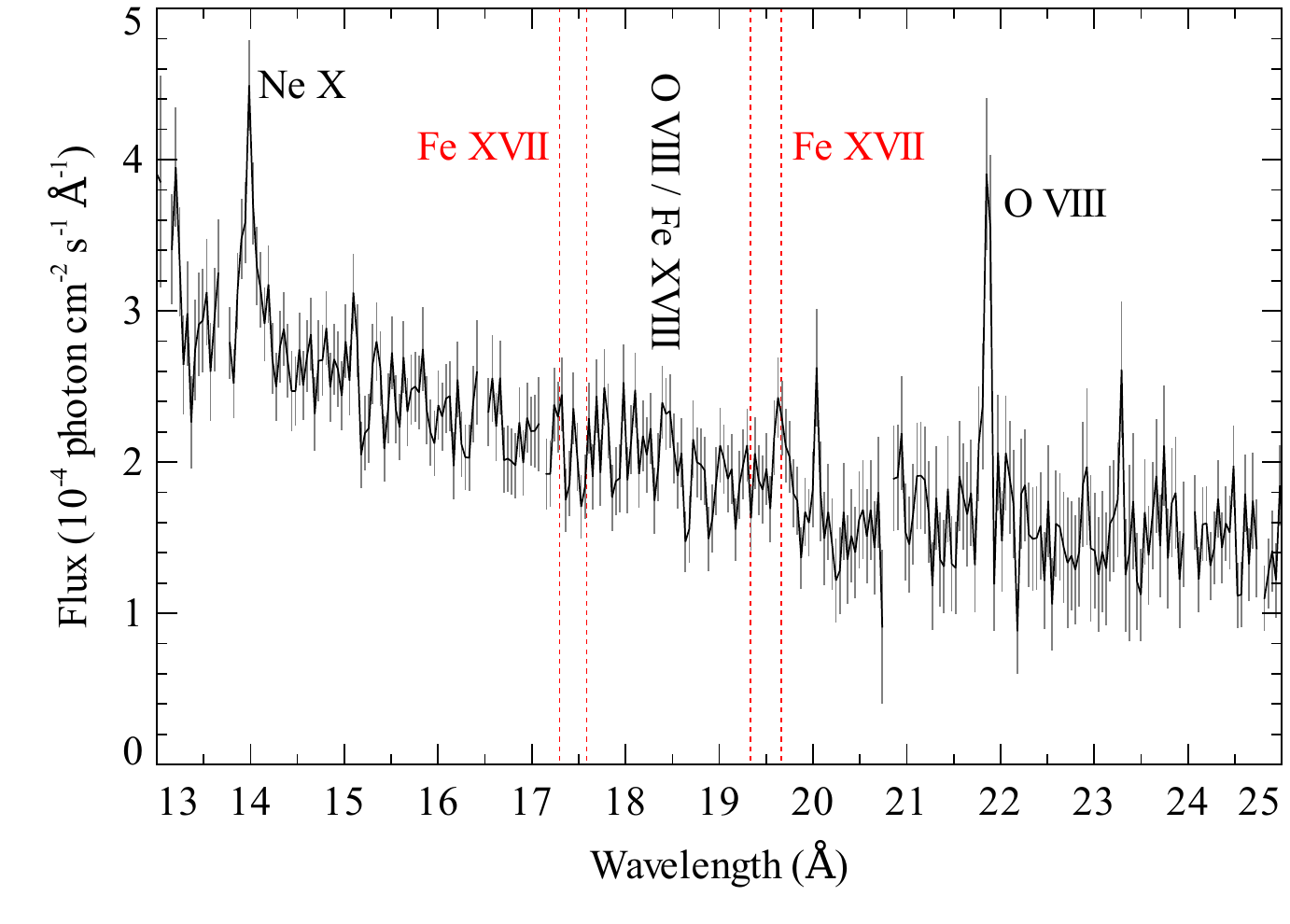}
  \caption{Fluxed and combined RGS spectra. This was fluxed using the
    \textsc{rgsfluxer} tool with 850 wavelength bins. Note that the
    spectral fitting was applied to the raw spectra, not this fluxed
    spectrum. The vertical dotted lines show the expected position of
    the Fe~\textsc{xvii} lines. The outer two lines should be the
    strongest ones.}
  \label{fig:rgsspec}
\end{figure}

We show in Fig.~\ref{fig:rgsspec} the fluxed version of the
spectrum. There are hints of the two strongest Fe~\textsc{xvii}
emission lines. Fitting zero-width Gaussian Fe~\textsc{xvii} lines to
the raw spectrum, and using an F-test to test for the significance of
the line components gives a chance probability of $1.5 \times 10^{-3}$
($3.2\sigma$) for the redshifted 17.06~{\AA} line, but only a chance
probability of 0.33 ($1\sigma$) for the redshifted 15.01~{\AA}
line. The 17.06~{\AA} line has a luminosity of $(7.6 \pm 2.6) \times
10^{41} \ergps$ (which is compatible with a $120\Msunpyr$ cooling
flow)..

Fitting the RGS data with a model made up of a thermal \textsc{vapec}
component and a \textsc{vmcflow} component cooling to 0.0808~keV gives
a mass deposition rate of $124\pm25 \Msunpyr$ (fixing the upper
temperature of the cooling flow to the \textsc{vapec} and using the
same free metallicities as above).

\section{Discussion}
\subsection{Inner surface brightness depressions}
The core of the cluster contains at least 5 or 7 X-ray surface
brightness depressions. If these are interpreted as X-ray cavities
caused by bubbles of relativistic plasma, then their $PV$ energies are
of the order of $8 \times 10^{57}$~erg (or a few times more for the
larger depression), where $P$ is the pressure of the surrounding ICM
and $V$ is the bubble volume. These are fairly typical values for
cavities in clusters \citep{DunnFabian04}. At the moderate distance of
this cluster, the depressions are hard to spatially resolve. We
observe no radio emission associated with the depressions, so we
cannot confirm them as radio bubbles. If they are radio cavities they
are ghost bubbles where the electrons have aged out of the observed
band.

\begin{figure}
  \centering
  \includegraphics[width=0.8\columnwidth]{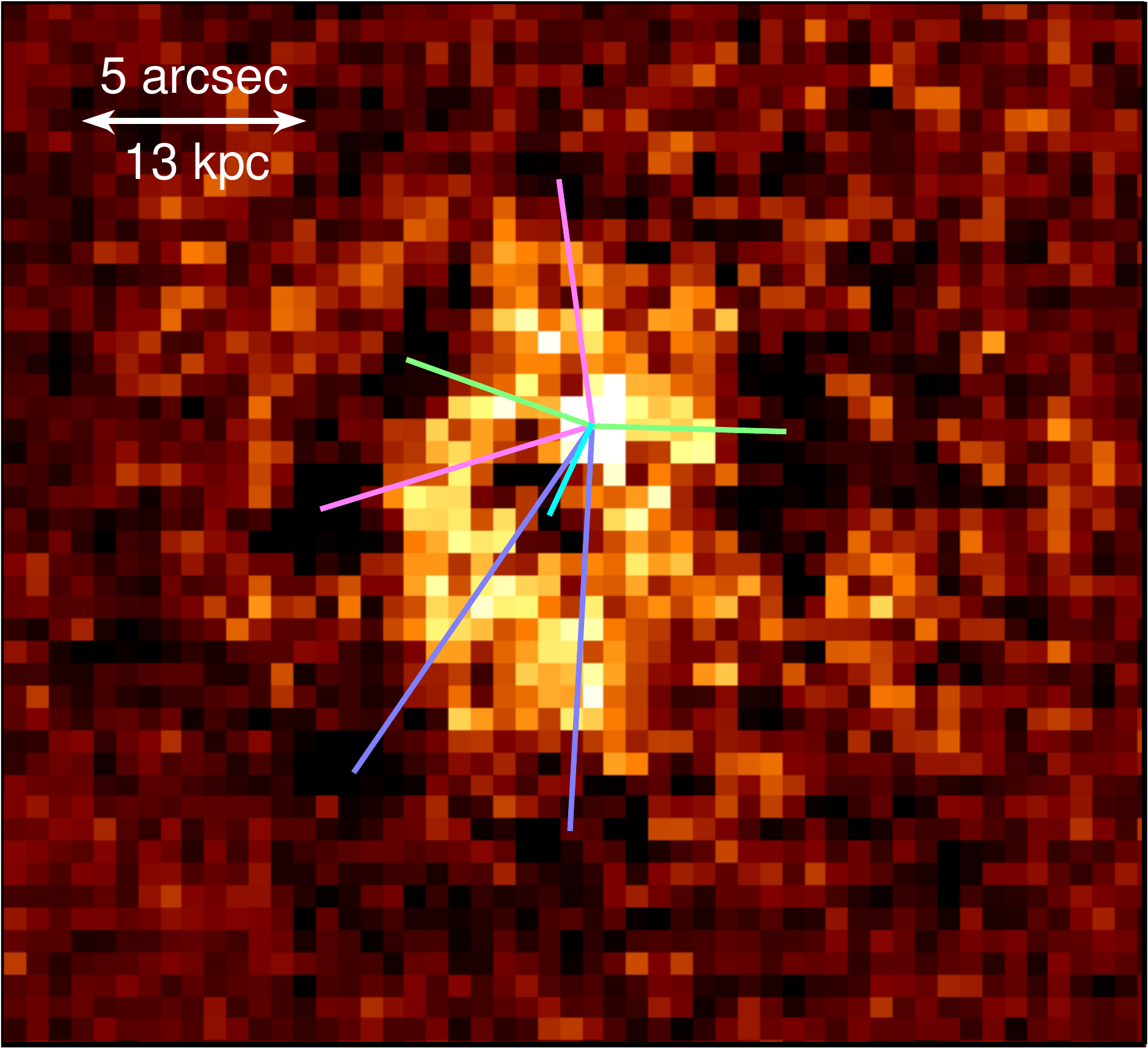}
  \caption{Unsharp-masked image of the very centre of the cluster,
    with lines connecting the radio/X-ray nucleus with the depressions
    in surface brightness. The image was created by subtracting the
    surface brightness image smoothed by Gaussians of $\sigma=0.5$ and
    5 pixels.}
  \label{fig:bubble_dia}
\end{figure}

Their morphology is rather unusual for a radio bubble, however. There
are 5 strong depressions, plus another two possible ones. They all lie
roughly around 20~kpc distance from the X-ray centroid. However, if
you measure the distance from the X-ray nucleus
(Fig.~\ref{fig:bubble_dia}), the north-east and west depressions at
similar radius ($\sim 12$~kpc), so are the north and east features
($\sim 16$~kpc), and the south and south-east regions ($\sim 25$~kpc),
leaving the central depression at 6~kpc distance. This may be
coincidental, but it fits with the picture of the bubbles being
created in in pairs (except for the central depression). We do not
know the component of distance from nucleus along the line of sight,
however. It could be that these are bubbles like the series of
`frothy' bubbles in M87 \citep{FormanM8707}.

Using a radius of 25~kpc (the maximum) and the sound speed implies a
timescale of $2.7 \times 10^7$~yr. Using $4PV$ enthalpy (if the gas in
the bubbles is relativistic; \citealt{Birzan04};
\citealt{DunnFabian04}), this would translate into a mechanical
heating rate of $4 \times 10^{43} \ergps$ per bubble.

The bolometric luminosity from the inner 100~kpc is $1.6\times 10^{45}
\ergps$. 5 bubbles would fall short of providing the required heating
rate to compensate cooling by an order of magnitude.

From the cavity heating power correlation of \cite{Birzan08}, the
1.4~GHz radio luminosity implies a cavity heating power of around
$10^{44} \ergps$. This power is similar to the calculated heating
power of the cavities. It still falls short of the required heating
power to prevent cooling by an order of magnitude. However, there are
a couple of orders of magnitude scatter in the correlation between
radio and heating power, so this is not at all conclusive.

We note that projection of cool gas in front of bubbles could create
the complex morphology observed. The bright emission to the east of
the bubbles in Perseus \citep{FabianPer00} is much stronger than the
deficit in the bubbles. If the bubbles were observed along a different
line of sight they may be completely obscured or difficult to
interpret. Therefore there may be much larger bubbles present in
Abell~2204 than we infer. We would require a bubble around the same
size as the bright central core to completely prevent cooling.

\subsection{Outer surface brightness depressions}
The cluster contains a surface brightness depression to the north,
between radii of 135 and 295~arcsec (355 to 780~kpc). To the south,
there is a depression between radii of 65 and 185~arcsec (170 to
490~kpc). Given their morphology, they are either old radio bubbles or
the cumulative result of many generations of radio bubble.

We can estimate how much energy each cavity could inject mechanically
into its surroundings as the enthalpy $4PV$. The average electron
pressure (Fig.~\ref{fig:profiles}) can be fitted outside 200~kpc
radius by a model with the form $P_\mathrm{e} = 0.156 [1 + (r/149.5
\kpc)^2]^{-1.13} \keVpcmcu$. If the total thermal pressure at each
radius is integrated over the volume of each bubble (assuming that
they are spherical), this leads to enthalpies for the northern and
southern cavities, respectively, of $9.9 \times 10^{61}$ and $1.3
\times 10^{62}$~erg (simply using $4PV$ with the pressure at the
midpoint gives a very similar result).  The total enthalpy is
therefore around $2 \times 10^{62} \erg$.

If they are a single set of bubbles, to estimate their mechanical
heating power, we need an appropriate timescale for the heating. If
the bubble is rising at a terminal velocity \citep{Churazov01}, then
the timescale for it to rise to its current radius is $t_\mathrm{buoy}
\sim R \sqrt{SC/2gV}$ \citep{Birzan04,DunnFabian04}, where $S$ is the
cross-sectional area of the bubble, $V$ is its volume, $R$ is the
distance of the bubble from the cluster nucleus, $g$ is gravitational
acceleration there and $C$ is a drag coefficient, 0.75.

For the northern bubble, at a radius of 220~arcsec in the cluster, the
mass enclosed is around $4.5 \times 10^{14} \Msun$
\citep{CloweSchneider02}, which with our assumed value of $H_0$,
implies $g \sim 1.9 \times 10^{-8} \cmpssq$.  This leads to a buoyancy
timescale of $2.7 \times 10^8 \yr$. The enclosed mass at the radius of
the southern bubble is $2.6 \times 10^{14}$, implying the buoyancy
timescale is $1.2 \times 10^8 \yr$. The total mechanical power of the
two bubbles is around $5 \times 10^{46} \ergps$. This value is larger
still than the most powerful outburst known, MS~0735+7421.

If these cavities are produced by a single episode, then it must have
avoiding heating the core of the cluster. We measure a minimum central
mean radiative cooling time of $2.5 \times 10^8$~yr.

If, however, the features are caused by the cumulative effect of many
generations of radio bubbles then we can estimate how long the
enthalpy would combat cooling in the cluster. The bolometric X-ray
luminosity within 500~kpc is calculated from our spectral fitting
results (Fig.~\ref{fig:profiles}) to be $3.5 \times 10^{45}
\ergps$. This means that the two features contain enough energy to
stop cooling for around 2~Gyr.

\subsection{Metallicity substructure}
Abell~2204 presents metallicity substructure from the innermost
regions (Fig.~\ref{fig:zoommaps} left panel) to several hundred kpc
radius (Fig.~\ref{fig:outermaps} bottom panel).

The metallicity substructure in the centre does not have any obvious
correlation with the observed X-ray cavities. Bubbles are expected to
drag cool, metal rich gas behind them \citep{Churazov01}. However,
when looking at the best data the picture is complex
\citep{SandersPer04,SandersPer07}, with some high metallicity regions
correlated with some bubbles. The metallicity structure on small
scales points towards the intracluster medium not being well mixed
\citep{SandersCent02,SandersPer04,SandersNonTherm05,Durret05,
  OSullivan05, Fabian05, Finoguenov06,SandersPer07, Simionescu08}.

The high metallicity blob in the central region
(Fig.~\ref{fig:zoommaps} centre panel; Fig.~\ref{fig:highzcentre}) has
a metallicity $1.2\Zsun$ greater than the neighbouring low metallicity
region (or four times its value). Using a volume of
$24\times12\times12$~kpc and an electron density of $0.1 \pcmcu$, this
represents an enhancement of iron of $2 \times 10^7 \Msun$.

\begin{figure}
  \centering
  \includegraphics[width=0.8\columnwidth]{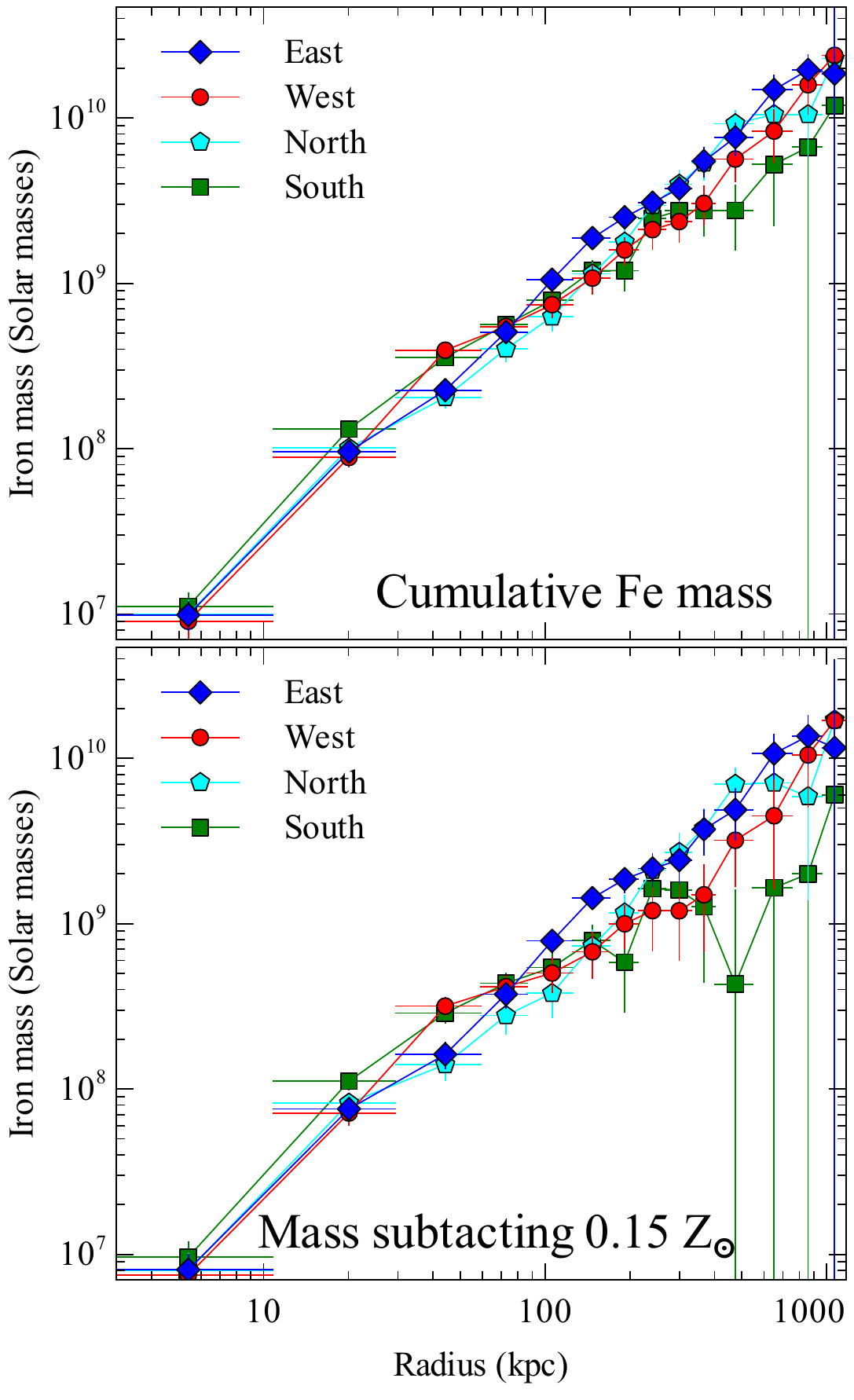}
  \caption{Cumulative iron mass profiles in the four different
    quadrants. The top panel shows the total Fe mass. The second shows
    the mass above a metallicity value of $0.15\Zsun$. The
    uncertainties were calculated using a Monte Carlo technique.}
  \label{fig:cumlfemass}
\end{figure}

To examine the outer northern metal enhancement, we calculated
cumulative mass profiles in the four different quadrants, shown in
Fig.~\ref{fig:cumlfemass} (also shown is the iron mass profile above a
metallicity value of 0.15, see \citealt{Bohringer04}). These results
assume the deprojected densities of Fig.~\ref{fig:quadne} and
projected metallicities of Fig.~\ref{fig:quadZ}. The uncertainties
were calculated by using a Monte Carlo calculation. Metallicity
profiles were simulated from the original profiles and measurement
errors (allowing them to be negative, using the larger side of the
error bar and assuming a Gaussian distribution). Cumulative mass
profiles were calculated from each simulated metallicity profile.  The
median, 16th and 84th percentiles were calculated to obtain the final
cumulative mass and uncertainties. The plot shows there is an excess
of a few times $10^9 \Msun$ of iron towards the north and east
compared to the south (the east is enhanced in mass compared to the
north as the density is higher there as is no surface brightness
depression). We note that metal masses can depend on the inhomogeneity
of the metallicity distribution \citep{Kapferer07}.

The northern outer metal enhancement may be associated with the cavity
to the north, as the high metallicity region is roughly at the base of
the northern cavity (Fig.~\ref{fig:outermaps}). The data quality is
not good enough to map this exactly. An outburst, a merger, or the
cumulative effect of many radio bubbles may be required to lift the
few times $10^9 \Msun$ of iron from the cluster core to the
outskirts. The metals could have been pulled out of the cluster core
in the wake of a rising bubble or many bubbles. In addition the
progress of the bubble(s) may have disturbed the cluster core enough
to make the inner metallicity spiral, though smaller outbursts may
have been responsible for this.

We note that the injection of metals into the intracluster medium
should be a smooth process. The Type Ia supernovae in the central
galaxy should be mainly responsible for the enrichment and these will
be uniformly distributed over the galaxy. Some other process must be
responsible for making the metallicity clumpy.

\subsection{Diffusion coefficient}
Given we have a high metallicity feature of the order of 6~kpc in size
and if we assume it lasts for $\sim 10^9$~yr, this gives a diffusion
coefficient of $< 10^{28} \cm^2 \s^{-1}$. This is similar to the value
obtained for Centaurus \citep{Graham06} and an order of magnitude
smaller than Perseus \citep{RebuscoDiff05}. Multiplying the sound
speed and the 6~kpc scale gives a diffusion coefficient around
$10^{30} \cm^2 \s^{-1}$. Unless these metallicity features are very
short lived, then diffusion must be heavily suppressed, implying
turbulent motions are damped on these scales.

\subsection{Cool gas}
The mass deposition rate calculated assuming that the X-ray luminosity
comes from cooling alone and is in steady state is around
$850\Msunpyr$ \citep{Peres98}. The RGS spectra from the cluster show
evidence for cool gas down to around 0.5~keV from spectral fitting,
and directly from the presence of the Fe~\textsc{xvii} line. The
quantity of cool gas is consistent with a cooling flow of $65
\Msunpyr$ down to the lowest detectable temperature. There is a wide
range of temperature of the X-ray emitting material in cluster from
around 12~keV all the way down to 0.5 keV. The gas below this
temperature is consistent to $2\sigma$ with $65 \Msunpyr$.

This level of cooling is also consistent with the star formation rate
of $14.7\Msunpyr$ measured from infrared emission \citep{Odea08}. We
note that the optical spectra of \cite{CrawfordBCS99} indicate only a
rate of $1.29\Msunpyr$. Feedback must be affecting the gas below
0.5~keV to prevent at least 80~percent of it cooling.

\section{Conclusions}
We observe large and significant surface brightness depressions 570
and 330~kpc from the core of the cluster of galaxies
Abell~2204. Morphologically they look like much larger versions of the
X-ray cavities seen in the cores of galaxy clusters, but energetically
they would be extremely powerful sources of mechanical heating if that
were the case ($5 \times 10^{46} \ergps$). They could, instead, be the
accumulation of multiple episodes of radio bubble formation in the
cluster core. The bubbles could rise in the same direction, forming a
bubble repository at large radius. If this is the case they would have
had to have avoided pancaking. The energy in this case could have
offset cooling in the cluster over 2~Gyr. The core of the cluster
contains several cavities, showing evidence for continued AGN
feedback.

We see a high degree of metallicity substructure in the intracluster
medium, from a 12~kpc feature containing $10^7 \Msun$ of iron in the
core, to a massive feature with $10^9 \Msun$ of iron at 400~kpc radius
to the north. These results, with other observations of the
intracluster medium, indicate that metals in the intracluster medium
are not efficiently mixed. The northern metallicity feature could have
been lifted by the giant outburst in the past, or the continuous
action of smaller outbursts.

In the core of the cluster are small depressions which may be cavities
generated by the nucleus. They could not provide enough heating to
prevent cooling in the central region, however. There is evidence for
cool X-ray emitting gas a factor of more than 10 lower in temperature
than the outer parts of the cluster. A cooling flow of around $65
\Msunpyr$ could be operating, in the absence of heating.

In the future, upcoming bolometers and other high energy resolution
detectors will allow us to gain a much better understanding of the
dynamical state of the ICM. We will be able to see the direct effect
of bubbles on the motion of the intracluster medium (although
filaments allow us to observe this indirectly, see
\citealt{FabianPerFilament03}) and examine the turbulence on different
scales.

\section*{Acknowledgements}
ACF thanks the Royal Society for support. GBT acknowledges support for
this work from the National Aeronautics and Space Administration
through Chandra Award Number GO7-8124X issued by the Chandra X-ray
Observatory Center, which is operated by the Smithsonian Astrophysical
Observatory for an on behalf of the National Aeronautics and Space
Administration under contract NAS8-03060.

\bibliographystyle{mnras}
\bibliography{refs}

\clearpage
\end{document}